\documentclass[aps,twocolumn,showpacs,superscriptaddress,floatfix]{revtex4-1}
\usepackage{graphicx,amsfonts,amssymb,amsmath,hyperref}

\newif\ifhyper
\hypertrue
\ifhyper
\hypersetup{
   citecolor = {red},
   colorlinks = {true}, 
   linkcolor = {blue},
   urlcolor = {blue} 
}
\fi

\def\be{\begin{equation}}
\def\ee{\end{equation}}
\def\bea{\begin{eqnarray}}
\def\eea{\end{eqnarray}}

 \newcommand{\Zd}{\mathbb{Z}_2}

 \newcommand{\MES}{\Xi}

 \newcommand{\ket}[1]{|#1\rangle}
  \newcommand{\bigket}[1]{\left|#1\right\rangle}
 \newcommand{\bra}[1]{\langle #1|}
 \newcommand{\braket}[2]{\langle #1|#2\rangle}

\newcommand{\id}{1\hspace{-.25em}{\rm I}} %
\newcommand{\Sm}{{\mathcal S}} %
\newcommand{\Um}{{\mathcal U}} %

\begin{document}

\title{Full characterization of a spin liquid phase: from topological entropy to robustness and braid statistics}



\author{Saeed S. Jahromi}
\email{jahromi@physics.sharif.edu}
\affiliation{Department of Physics, Sharif University of Technology, Tehran 14588-89694, Iran}

\author{Abdollah Langari}
\email{langari@sharif.edu}
\affiliation{Department of Physics, Sharif University of Technology, Tehran 14588-89694, Iran}
\affiliation{Center of Excellence in Complex Systems and Condensed Matter, Sharif University of Technology, Tehran 14588-89694, Iran}

\begin{abstract}
We use the topological entanglement entropy (TEE) as an efficient tool to fully characterize 
the Abelian phase of a $\Zd \times \Zd$ spin liquid emerging as 
the ground state of topological color code (TCC), which is a class of stabilizer states on the honeycomb lattice.
We provide the fusion rules of the quasiparticle (QP) excitations of the model by introducing single- or two-body 
operators on physical spins for each fusion process which justify the corresponding fusion outcome.
Beside, we extract the TEE from Renyi entanglement entropy (EE) of the TCC, analytically and numerically 
by finite size exact diagonalization on the disk shape regions with contractible boundaries.
We obtain that the EE has a local contribution, which scales linearly with the boundary length 
in addition to a topological term, i.e. the TEE, arising from the condensation of 
closed strings in the ground state. We further investigate the ground state dependence of the TEE on regions with non-contractible
boundaries, i.e. by cutting the torus to half cylinders, from which we further identify multiple 
independent minimum entropy states (MES) of the TCC and then extract the $\Um$ and $\Sm$ 
modular matrices of the system, which contain the self and mutual statistics of the anyonic QPs and fully characterize the 
topological phase of the TCC.
Eventually, we show that, in spite of the lack of a local order parameter, TEE and other physical quantities obtained 
from ground state wave function such as entanglement spectrum (ES) and ground state fidelity are sensitive probes to study the robustness of a 
topological phase. We find that the topological order in the presence of a magnetic field persists until the vicinity of the transition point, where the TEE and fidelity drops
to zero and the ES splits severely, signaling breakdown of the topological phase of the TCC.

\end{abstract}
\maketitle

%
%
\section{Introduction}
%
%
Characterizing microscopic features of a Hamiltonian and its order has always been a cumbersome task particularly for the 
so called {\it topologically ordered} phases of matter, \cite{wen0} which lie beyond the Ginzburg-Landau paradigm \cite{landau}. Due to the lack of a local order parameter,
detection of topological order (TO) in the ground state of a microscopic Hamiltonian is a very difficult task. Furthermore, distinguishing between distinct classes of TO 
observed in different quantum systems such as the fractional quantum Hall systems \cite{tsui}, high-temperature superconductors \cite{wen1,superconductor1,superconductor2}, 
highly frustrated magnetic systems \cite{RVB1,RVB2,RVB3,RVB4} and recently in the context of topological quantum computation \cite{Kitaev1}, has already been a challenging mission.

Ground states of the topologically ordered phases have long-range entanglement \cite{chen} which lead to many interesting features such as topological degeneracy and the 
fractionalized emergent quasiparticles (QP) with anyonic statistics \cite{Kitaev1,wen2,wen3}. 
The long-range entanglement in the ground states of 
the topological phases can furtherer be utilized for fault tolerant quantum computation by defining non-local quantum bits on the topological degrees 
of freedom to protect the information from local decoherence \cite{Kitaev1,Kitaev2}.
In spite of the lack of local order parameter, the entanglement can be used as a reliable source for identifying and characterizing the TO in the ground state of a system 
by resorting to the concept of topological entanglement entropy \cite{K-P-TEE,L-W-TEE}. 

The statistics of the quasiparticle of the system can further be extracted from the TEE. The braid statistics of the anyonic excitations is given by 
the modular matrices $\Um$ and $\Sm$ which provide details about the self and mutual statistics of the QPs, respectively \cite{superconductor1,nayak,Keski,Francesco,kitaev3}.
Distinct features of the underlying topological order such as the fusion rules of the QPs and total quantum dimension of the anyonic model are further perceived from the 
elements of the $\Sm$-Matrix \cite{Verlinde}. The modular matrices can be extracted from the knowledge of entanglement in the ground state of a topological phase and the full 
characterization of the TO is therefore possible \cite{Haldane1,Haldane2,Grover,Pollmann,Vishwanath,Vidal}. 

Topological color code \cite{bombin_distilation} is an example of topologically ordered system, which its ground state describes a spin 
liquid phase with $\Zd \times \Zd$ topological order. Due to its special structure on the trivalent lattices and an interplay between color 
and homology in the construction of the code, the number of encoded logical qubits in 
the color code is twice the number of Kitaev's toric code \cite{Kitaev1}. 
This allows the full implementation of the whole Clifford group in a fully topological manner in the ground state space and make the fault tolerant quantum 
computation possible, without resorting to the braiding of QPs. 
It is possible to show that the color code is equivalent to two copies of the toric code and therefore the two models belong to the same family of the universal topological
phases \cite{tcc_tc1, tcc_tc2, tcc_tc3}.  
Recently, a minimal version of the color codes, namely the triangular codes, has been realized experimentally which is a step forward in building 
a quantum memory based on the topological color codes \cite{tcc_experimental}. We are therefore motivated to study the topological phase of the model in more depth
and fully characterize the Abelian phase of the system.

Entanglement properties of the TCC on different bipartitions of the honeycomb lattice with contractible boundaries has already been studied 
by counting the number of colored strings in each subregions \cite{kargarian_entaglement}. In this paper, we use a different approach and 
study the topological entanglement entropy of the system both on regions with contractible and non-contractible regions 
from Renyi entanglement entropy. On a disk-shape region with contractible boundaries, we calculate the Renyi entropy 
analytically and numerically by finite size exact diagonalization on honeycomb
clusters with different sizes. We show that the EE has a local contribution, 
which scales linearly with the boundary length (that is a manifestation of the 
so called {\it area} law \cite{area_law1,area_law2,area_law3}) and a universal term, 
which has a topological nature, i.e. the TEE, stemming from the condensation of closed strings in the ground state. 

We further investigate the ground state dependence of the TEE by calculating the Renyi entropy on regions with non-contractible 
boundaries, i.e. by cutting the torus to half cylinders, and identify the multiple independent 
minimum entropy states \cite{Vishwanath} of the TCC, which form a set of orthogonal basis states for the ground state manifold. We show that the MESs are the 
simultaneous eigenstates of the Wilson loop operators and characterize the MESs by defining certain types of loop insertion operator, which relate each MES to 
a distinct QP. We further extract the $\Um$ and $\Sm$ modular matrices of the TCC from MESs and fully characterize the TO in TCC.

Eventually, we study the stability of TO in the ground state of TCC in the 
presence of a magnetic perturbation by means of topological entanglement entropy, entanglement spectrum and ground-state 
fidelity \cite{fidelity}. 
Although, robustness of the TCC in the presence of a single parallel magnetic field has already been examined by one of the authors 
through analyzing the energy spectrum of the system with high-order series expansion technique \cite{ssj,ssj2,ssj3}, we believe that 
observing the topologically ordered ground state, 
while the perturbation is tuned, will provide a more comprehensive picture about 
the robustness and its corresponding quantum phase transition. 
Our finding further reveals that the transition point obtained by TEE, ES and the ground-state fidelity
confirm the corresponding values detected by analyzing the low-energy spectrum of the system \cite{ssj}.

The outline of our paper is as follows. In Sec.~\ref{sec:TCC}, we briefly review the TCC model and some of its important features. 
We extract the fusion rules of the QPs and their physical roots in Sec.~\ref{sec:fusion}.
The topological entanglement entropy of the system on disk shape regions is calculated both analytically and 
numerically in Sec.~\ref{sec:TEE}.
In Sec.~\ref{sec:GRD-TEE}, we envisage the ground state dependence of the TEE  by cutting the torus to half cylinders. Next, we   
calculate the MESs of the TCC in Sec.~\ref{sec:MES}. Thereafter in Sec.~\ref{sec:Braid-TEE}, we extract the $\Um$ and $\Sm$ modular 
matrices of the TCC from MESs and fully characterize the TO of TCC.
In section \ref{sec:QPT-TEE}, we probe the topologically ordered 
ground state, while tuning a parallel magnetic 
field in the $x$-direction and calculate TEE, ES and fidelity to estimate the stability of the TO in the ground state of the TCC. 
We further provide a more clear picture about the breakdown of the topological phase in this section.
Finally, Sec.~\ref{sec:conclude} is devoted to the conclusion.

\section{Topological Color Code}
\label{sec:TCC}

\begin{figure}
\centerline{\includegraphics[width=7cm]{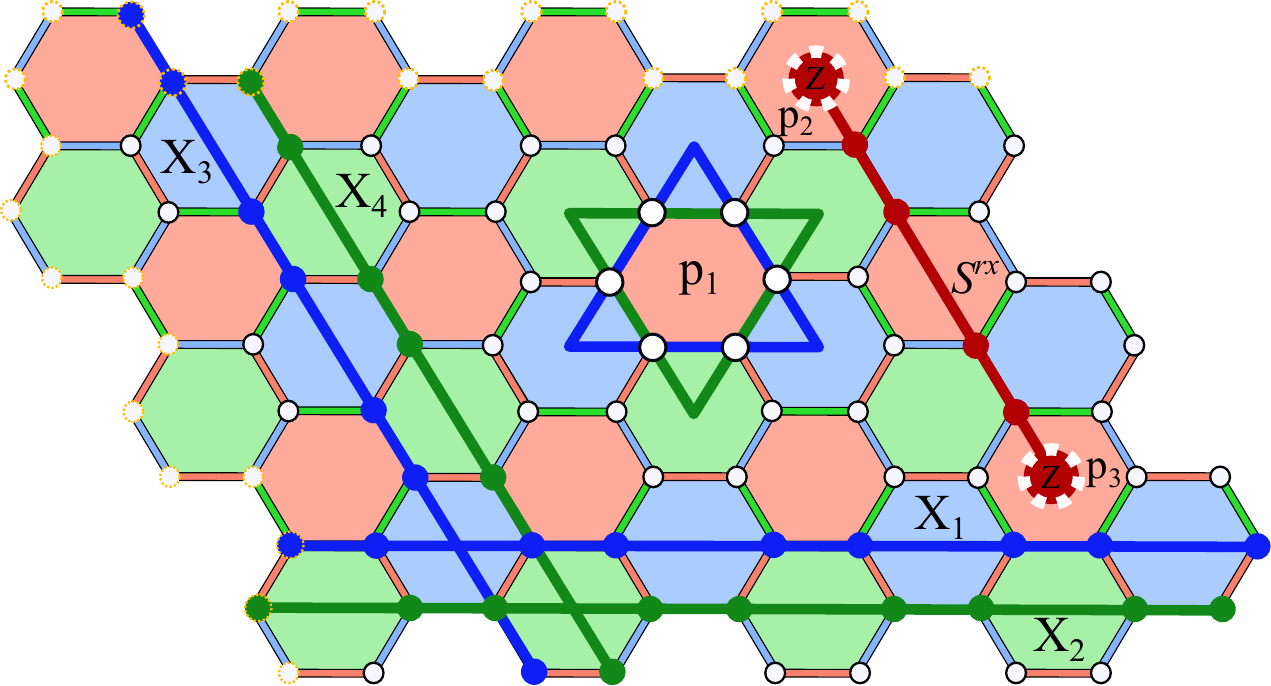}}
\caption{(Color online) TCC on the honeycomb lattice $\Lambda$ placed on a torus with genus $g=1$.
A red plaquette $p_1$ is characterized by two red and green closed strings at its boundary. An open red string ($S^{rx}$)
creates two $Z$-type QPs at its end points on the surface of the two red plaquettes $p_2$ and $p_3$.
For every homology class of the torus, there are four global strings which can be labeled as ($X_1$, $\ldots$, $X_4$).
The global strings are responsible for the $16$-fold topological degeneracy of the ground state.}
\label{fig:tcc}
\end{figure}

In this section, we recall the notions of topological color code. Consider a 2D colorable and trivalent lattice which is a collection of vertices, links and faces (plaquettes), embedded on an arbitrary manifold of genus $g$.
Each vertex of the lattice is connected to three links and each link connects two plaquettes of 
the same color and shares the same color with the plaquettes.
Such a structure is called a 2-colex, \cite{bombin_distilation, bombin_statistcal} 
which is illustrated in Fig.~\ref{fig:tcc} as an example on a torus with $g=1$.

To each plaquette $p$, we associate two distinct operators which are the product of Pauli spins on the vertices of the plaquette and are defined as 
$X_p=\prod_{i \in p} \sigma_i^x$ and $Z_p=\prod_{i \in p} \sigma_i^z$. These operators can be classified according to their color to three sets $r$, $g$ and $b$.
It is straightforward to check that 
\bea\label{eq:plaquette_superfluous}
\prod_{p\in r}X_p &=& \prod_{p\in g}X_p = \prod_{p\in b}X_p, \\
\prod_{p\in r}Z_p &=& \prod_{p\in g}Z_p = \prod_{p\in b}Z_p, 
\eea
which implies four of the generators are superfluous \cite{bombin_distilation}. 
The Hamiltonian of the topological color code on a trivalent lattice $\Lambda$ is given by the sum over all plaquette operators i.e. \cite{bombin_distilation} 
\be\label{eq:H_TCC}
H_{\rm TCC}=-J\sum_{p \in \Lambda} (X_p+Z_p).
\ee
All terms in the above Hamiltonian commute with each other, resulting in exact solubility of the model. The plaquette operators further satisfy the square-identity relation, 
$(X_p)^2=\id=(Z_p)^2$,  with eigenvalues $x_p=z_p=\pm1$. Choosing the $+1$ eigenvalues as desired ones and setting $J>0$, the ground-state energy of the 
model on a lattice with $N_p$ plaquette ($2N$ sites) reads $E_0=-2N_pJ$. 
Elementary excitations of the model are further gapped quasiparticles, which are local on the surface of the 
plaquettes and correspond to their $-1$ eigenvalues.

Ground state of the system is constructed from the product of the plaquette operators acting on a reference state and is given by:
\be\label{eq:grstate}
\ket{\xi_{0000}}=\frac{1}{\sqrt{|G|}} \sum_{g\in G} g \ket{0}^{\otimes N},
\ee
where $N$ is the number of lattice sites and $\ket{0}$ is the eigenstate of the $\sigma^z$ Pauli operator such that $\sigma^z\ket{0}=\ket{0}$.
Furthermore, $G$ is the group ($g\in G$) constructed by all possible products of the plaquette operators with spin flip capabilities i.e. the $X_p$ operators. 
The cardinality of the group, $|G|$, on a lattice with $N_p$ plaquettes is $2^{N_p-2}$ \cite{kargarian_entaglement} (this is the direct 
consequence of Eq.~(\ref{eq:plaquette_superfluous})).

Another spectacular feature of the TCC is the existence of strings, which are generalization of the plaquettes and are defined
as particular paths built by connecting a series of links with the same color and can be open or closed \cite{bombin_statistcal}. 
The corresponding string operators, similar to the plaquettes, are the product of Pauli spins on the string.
The ground state (\ref{eq:grstate}) is a uniform superposition of highly fluctuating closed strings, 
which is typical for systems with TO \cite{Kitaev1,wen3}, and for the case of TCC is denoted by a $\Zd\times\Zd$ spin liquid. 

Wrapping the system around a Riemannian surface, a special family of the closed strings emerge, which are non-contractible and responsible
for the topological degeneracy of the ground state of the system. These global loops for the color code is denoted by ($X_{1}, \ldots, X_{4}$) 
\cite{bombin_distilation, bombin_statistcal} and are illustrated in Fig.~\ref{fig:tcc}.
Applying the global strings to the ground state (\ref{eq:grstate}), yields the set of degenerate states which 
all together form the $16$-fold topologically degenerate ground space of the system \cite{bombin_distilation}
\be \label{eq:degenerate_gs}
{\cal C}=\{\ket{\xi_{ijkl}}: \ket{\xi_{ijkl}}=X_{1}^{i} X_{2}^{j} X_{3}^{k} X_{4}^{l} \ket{\xi_{0000}} \},
\ee 
where, $i,j,k,l=1(0)$ correspond to the global string $X$ appearing (not appearing) in the set. One can further write a generic form of the ground state
as an equal superposition of the states from the ground space as
\be\label{eq:general_grs}
\ket{{\bf \Psi}}=\sum_{i,j,k,l=0,1} a_{ijkl} \ket{\xi_{ijkl}}, \quad \sum_{i,j,k,l=0,1} |a_{ijkl}|^2=1.
\ee

\section{Emergent quasiparticles and their Fusion}
\label{sec:fusion}
\begin{figure}
\centerline{\includegraphics[width=7cm]{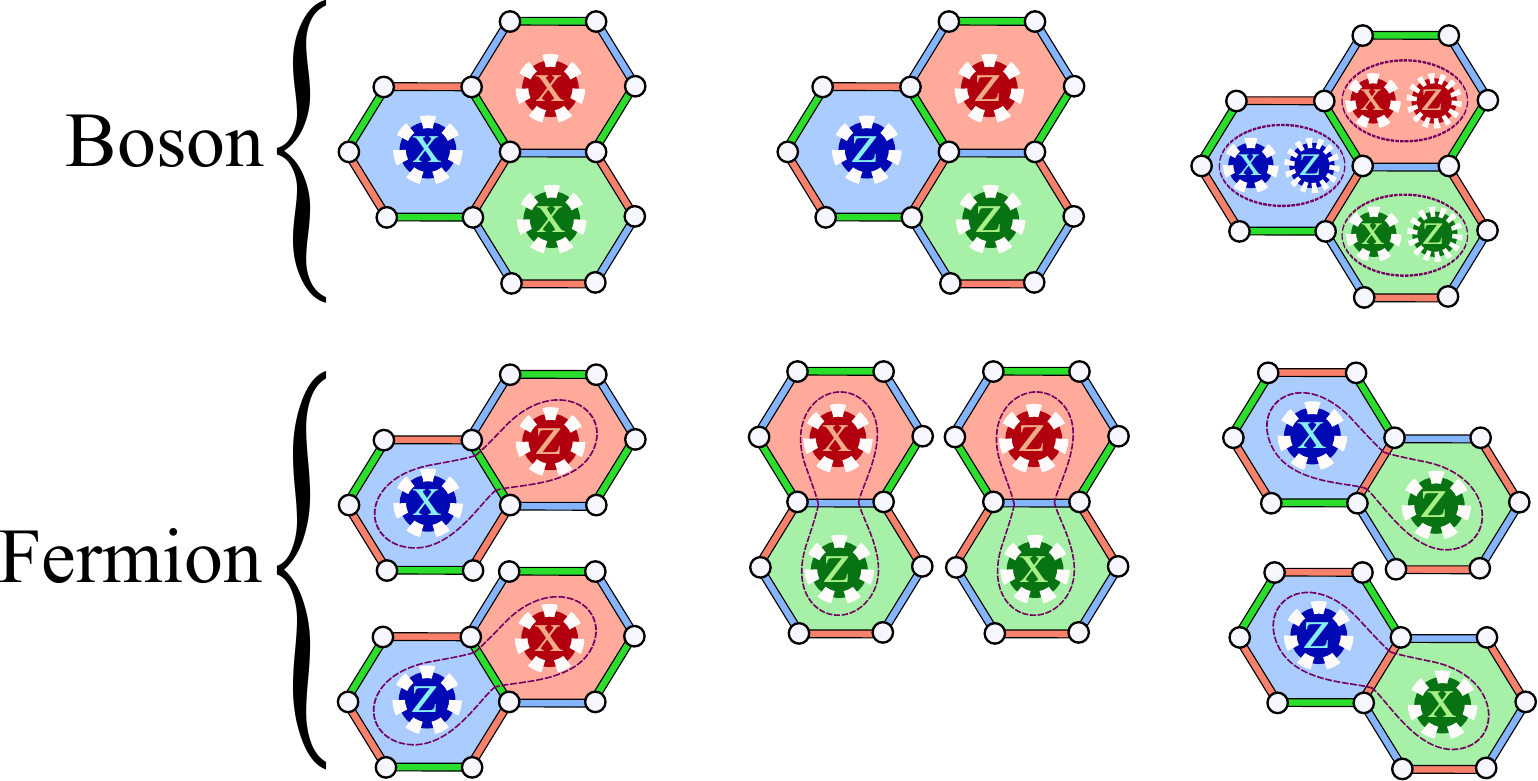}}
\caption{(Color online) The whole spectrum of the excitationas of the TCC. The elementary excitations as well as the combination of different excitations with the same 
color are boson. The combination of elementary QPs with different color and type forms two families of fermions.}
\label{fig:excitations}
\end{figure}

Elementary excitations of the TCC model correspond to the $-1$ eigenvalues of the plaquette operators which are localized at the end points of open strings 
and have the character of color \cite{bombin_anyonic_intract}. On a torus, the QPs are always created in pairs and one can be considered as the anti-particle
of the other QP. See for example the $Z$-type excitations on the surface of the red plaquettes $p_2$ and $p_3$ in Fig.~\ref{fig:tcc}.

The elementary excitations are Abelian anyons and excitations with different type and color have semionic mutual statistics i.e. 
the $-1$ phase they pick up as a result of braiding, corresponds to a $\pi$ phase which is one half of the phase one would get for fermions. 
Furthermore, based on the color and type of the elementary excitations, different emergent particles can be combined. 
The elementary excitations as well as the combination of different excitations with the same color are boson.
However, the combination of elementary QPs with different colors and types form
two families of fermions \cite{note1, bombin_two_body}. 

Each quasiparticle excitation has a character of color $c=r,g,b$ and carries a topological 
charge $q=e$,($m$) -- representing  $X$- ($Z$-)type QPs-- where $e$ and $m$ imitate the electric 
field and magnetic flux of a $\Zd$ gauge theory, respectively.
The particles are denoted in general by $\chi^c_q$.
However one should note that the color code belongs to the $\Zd \times \Zd$ gauge theory 
which is a consequence of an interplay between color and homology \cite{bombin_anyonic_intract}.
Considering the vacuum as a QP with trivial charge, TCC poses $16$ topological charges represented by
\be
\begin{gathered}\label{eq:qps}
 \{ {\bf 1} , \chi_e^r , \chi_e^g , \chi_e^b , \chi_m^r , \chi_m^g , \chi_m^b , \chi_{e}^{r} \chi_{m}^{r}, \chi_{e}^{g} \chi_{m}^{g} , \chi_{e}^{b} \chi_{m}^{b} \}, \\
\{ \chi_{e}^{r} \chi_{m}^{g} , \chi_{e}^{g} \chi_{m}^{r} , \chi_{e}^{r} \chi_{m}^{b} , \chi_{e}^{b} \chi_{m}^{r} , \chi_{e}^{b} \chi_{m}^{g} , \chi_{e}^{g} \chi_{m}^{b} \},
\end{gathered}
\ee
where the members of the first (second) set are bosons (fermions). The whole spectrum of the excitations is illustrated in Fig.~\ref{fig:excitations}.

Due to the special structure of the TCC model, any local operators on vertices of the lattice can excite the plaquette operators which share the vertex.
For example, the $\sigma_i^x$ operators anti-commute with the three $Z_p$ plaquette operators which are attached to site $i$ and three fluxes with
different color are created on the surface of the plaquettes.  Similar rules holds for $\sigma_i^z$ and $\sigma_i^y$ operators, as well. However, one should note that 
the $\sigma_i^y$ operators anti-commute with $X_p$ and $Z_p$ operators, simultaneously and creates six QPs at the same time. This can be the physical source
behind the fusion of quasiparticles. In what follows, we show by several examples that certain types of single- or two-spin operators can fuse the QPs and result in a 
distinct fusion outcome. We can then generalize these actions into simple rules and drive the fusion algebra of the Abelian QPs of the TCC. 

\begin{figure}
\centerline{\includegraphics[width=6cm]{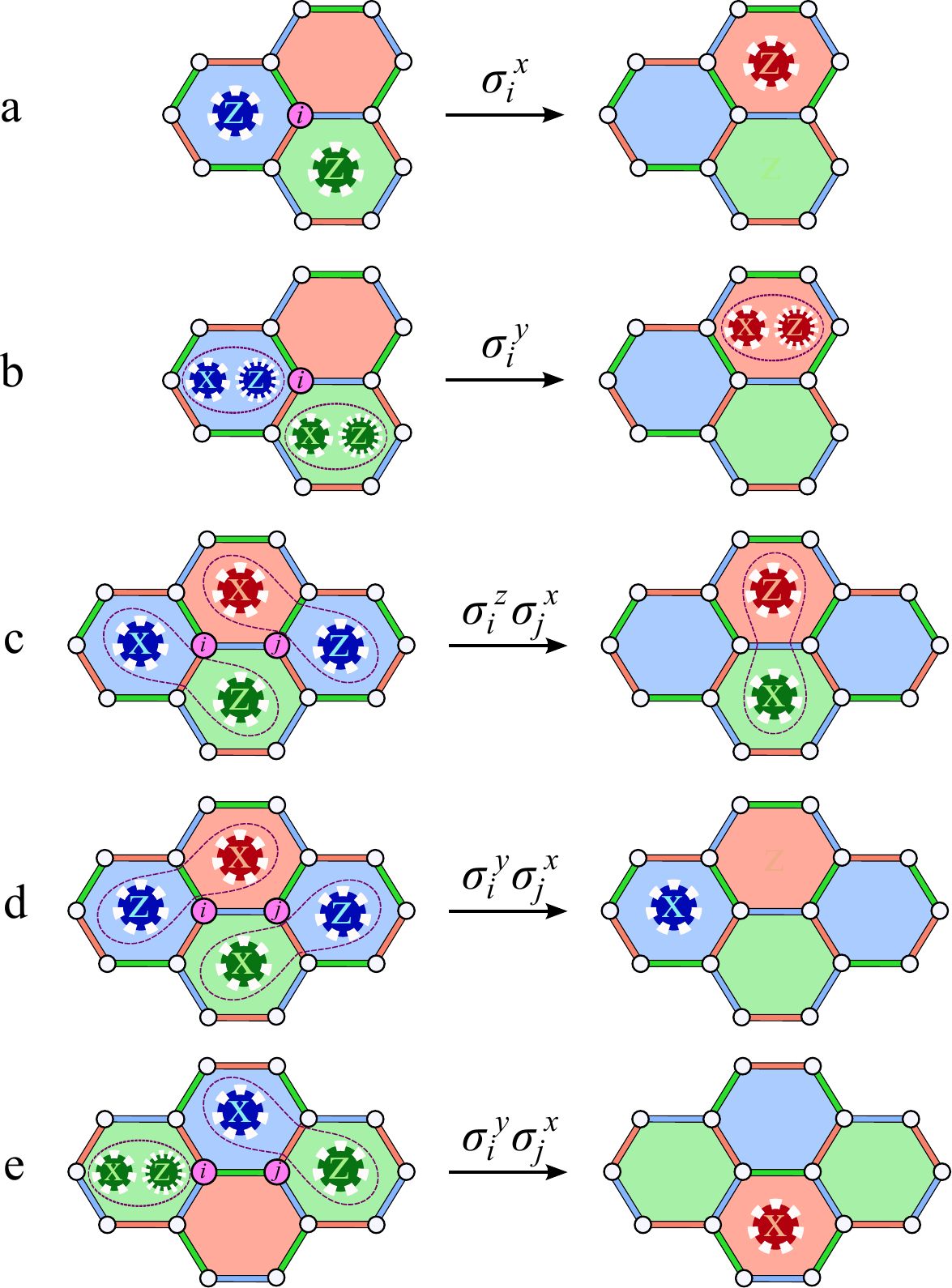}}
\caption{(Color online) Examples of quasiparticles fusion by single- and two-spin operations.}
\label{fig:fusion}
\end{figure}

For the first example, consider two $\chi_m^g$ and $\chi_m^b$ fluxes on the surface of a neighboring green and blue plaquettes (Fig.~\ref{fig:fusion}-a).
Action of $\sigma_i^x$ operator on site $i$ will annihilate the the two fluxes and create $\chi_m^r$ on the red plaquette:
\be
\chi_m^g \times \chi_m^b=\chi_m^r, \quad \quad \times :\equiv \sigma_i^x.
\ee
Next, let us fuse two composite bosons $\chi_e^b\chi_m^b$ and $\chi_e^g\chi_m^g$ under the action of $\sigma_i^y$ operator (Fig.~\ref{fig:fusion}-b). This will 
destroy the two green and blue bosons and create a red one on the surface of the neighboring red plaquette: 
\be
\chi_e^b\chi_m^b \times \chi_e^g\chi_m^g=\chi_e^r\chi_m^r, \quad \quad \times :\equiv \sigma_i^y.
\ee
Now, we fuse two fermions $\chi_e^b\chi_m^g$ and $\chi_e^r\chi_m^b$ with a two-body operation $\sigma_i^z \sigma_j^x$ (Fig.~\ref{fig:fusion}-c).
The resulting QP as a consequence of the on-site anti-commutation of the plaquette and Pauli operators is a $\chi_e^g\chi_m^r$ fermion:
\be
\chi_e^b\chi_m^g \times \chi_e^r\chi_m^b=\chi_e^g\chi_m^r, \quad \quad \times :\equiv \sigma_i^z \sigma_j^x.
\ee
Other fusion examples in Fig.~\ref{fig:fusion}-d,e similarly read
\bea
\chi_e^r\chi_m^b \times \chi_e^g\chi_m^b &=& \chi_e^b, \quad \quad \quad \times :\equiv \sigma_i^y \sigma_j^x, \\
\chi_e^g\chi_m^g \times \chi_e^b\chi_m^g &=& \chi_e^r, \quad \quad \quad \times :\equiv \sigma_i^y \sigma_j^x. 
\eea
As the QPs are Abelian, there exist only one fusion outcome for each fusion process. Therefore, among all possible 
single or two-spin operators, only those which result in a single fusion outcome are valid. Fusion of other QPs with identity particle is further equivalent to not acting with any local operator on
the lattice.

After identifying all single- and two-spin operators and their fusion outcome,
the results can be generalized in the following fusion rules:
\be\label{eq:fusion_rules}
\begin{gathered}
\chi^c_q \times {\bf 1}={\bf 1} \times \chi^c_q =\chi^c_q , \\
\chi^c_q \times \chi^{c'}_{q}=\delta_{cc'}+({\bf 1}-\delta_{cc'}) \times \chi^{c \star c'}_q ,\\
\chi^c_q \times \chi^{c'}_{q'}=\chi^c_q \chi^{c'}_{q'} ,
\end{gathered}
\ee
where ${\bf 1}$ denotes the vacuum QP and $\delta_{cc'}$ is the usual Kronecker delta. Defining the bar symbol as an operator that transforms colors cyclically as
$\bar{\rm r}=\rm g$, $\bar{\rm g}=\rm b$ and $\bar{\rm b}=\rm r$, the $\star$ which is a symmetric color operator is given by
\be\label{eq:color_star}
c \star c=c, \quad c \star \bar{c}=\bar{c} \star c= \bar{\bar{c}}.
\ee
Let us stress that the fusion outcome of a composite particle is determined by applying these rules to each of its components, individually.

Using these fusion rules, the fusion table of the TCC is provided in Table~\ref{tab:fusion_table} in appendix,
which contain all of the possible fusion processes and their outcome. One can check that the examples of Fig.~\ref{fig:fusion} are in 
correct correspondence with the fusion table.

\section{Topological Entanglement Entropy}
\label{sec:TEE}
Due to the lack of a local order parameter, detection of topologically ordered phases is usually a cumbersome task. Kitaev, Preskill \cite{K-P-TEE} and Levin, Wen \cite{L-W-TEE} 
independently showed that the topological order in the ground state of a topologically ordered phase can be detected by probing the entanglement entropy of 
a subregion of the system, as long as the size of the subregion is larger than the correlation length. 

In the following, we first analytically extract the TEE of color code by calculating the Renyi entropy of the system. 
Next, we calculate  the TEE numerically by implementing the Kitaev-Preskill (KP) strategy \cite{K-P-TEE} and 
show that there is a full correspondence between the analytical and numerical results. 
A brief review on basic concepts of the TEE and the KP strategy is further provided in appendix \ref{AP:EE}.

\subsection{Extracting the TEE from Renyi Entropy}
In order to calculate the Renyi entropy, we bipartition the lattice by considering a disk shape region with contractible boundary as subsystem $A$ and denote the 
rest of the latices as subsystem $B$. Considering the system to be in a generic state $\ket{\bf \Psi}$ which is defined in Eq.~\ref{eq:general_grs}, the reduced density matrix 
(RDM) of region $A$ is given by \cite{kargarian_entaglement,Hamma-TEE}
{\small
\be
\rho_A=\sum_{ijkl,mnpq} a_{ijkl}  a^{*}_{mnpq} {\rm Tr}_B (X_{1}^{i} X_{2}^{j} X_{3}^{k} X_{4}^{l} \rho_0 X_{1}^{m} X_{2}^{n} X_{3}^{p} X_{4}^{q}), 
\ee}
where $\rho_0=\ket{\xi_{0000}}\bra{\xi_{0000}}$ is the density matrix of the state with no global 
loop acting on the system.
Whenever, the subsystem $A$ has a disk shape geometry 
with contractible boundaries, one can show \cite{kargarian_entaglement} any global string passes through the region can be deformed out of the 
boundaries of $A$ and the RDM of the subsystem reads
%
%
\be\label{eq:TCC-RDM}
\rho_A={\rm Tr}_B (\rho_0).
\ee
Eq.(\ref{eq:TCC-RDM}) implies that only different configurations of contractible loops intersect the boundaries of the region. In other words, for a disk shape geometry 
only those configurations are allowed which cross the boundary even number of times. 

In order to count the number of these configurations, we introduce 
subgroups of $G$ acting only on the subsystems $A$ and respectively $B$ by 
\bea
G_A&=&\{g\in G|g=g_A\otimes \id_B \}, \nonumber \\
G_B&=&\{g\in G|g=\id_A\otimes g_B \}. 
\eea
Denoting the subgroup of $G$ which acts simultaneously on both subregions (boundary of the entanglement partition) by $G_{AB}$,
the number of loop configurations acting on the boundary of subsystem $A$ is given by the cardinality of the subgroup i.e. $|G_{AB}|=\frac{|G|}{|G_A||G_B|}$ 
where $|G_A|$, $|G_B|$ are orders of the subgroups. It is straightforward to show that for a disk 
shape geometry of the TCC,  $|G_{AB}|=2^{L-2}$, where $L$ is the length of 
the boundary \cite{kargarian_entaglement}.

The Schmidt decomposition of the ground state wave-function in regions $A$ and $B$ can therefore be indexed by $g \in G_{AB}$ i.e.
\be
\ket{\xi_{0000}} = \frac{1}{\sqrt{|G_{AB}|}}  \sum_{\{g\} \in \rm even} \ket{\Psi^A_{\{g\}}} \ket{\Psi^B_{\{g\}}}.
\ee

Using this Schmidt-decomposed ground state and Eq.(\ref{eq:TCC-RDM}), the RDM of region $A$ therefore reads
\be 
\rho_A=\frac{1}{|G_{AB}|} \sum_{\{q\} \in \rm even} \ket{\Psi^A_{\{q\}}} \bra{\Psi^A_{\{q\}}}.
\ee

From the above equation it is immediately followed that the Renyi entanglement entropy of the TCC for a region with contractible boundary is given by
\bea
S_n &=& \frac{1}{1-n} \log_2 \left((\frac{1}{|G_{AB}|})^{n} |G_{AB}| \right)\nonumber \\ 
&=& \frac{1}{1-n} \log_2 \left( (2^{L-2})^{-(n-1)} \right) \nonumber \\ 
&=& L \log_2 2 -2 \log_2 2 =L - \log_2 4. \label{eq:RenyTCC}
\eea
The first term in Eq.~(\ref{eq:RenyTCC}) shows that the Renyi EE for the topological 
color code scales linearly as $\alpha L$ with the boundary length $L$ of the 
subregion $A$ with $\alpha=1$, which is a manifestation of the {\it area} 
law \cite{area_law1,area_law2,area_law3}. The second term is further referred to as 
topological entanglement entropy, $\gamma=\log_2 \mathcal{D}$,
where for the TCC $\gamma=\log_2 4=2$. Let us stress that what we obtain for the 
TEE is independent of the Renyi index $n$ \cite{Dong,Flammia}, and is the expected value
for a $\Zd \times \Zd$ gauge theory with total quantum dimension $\mathcal{D}=4$.

\begin{figure}
\centerline{\includegraphics[width=\columnwidth]{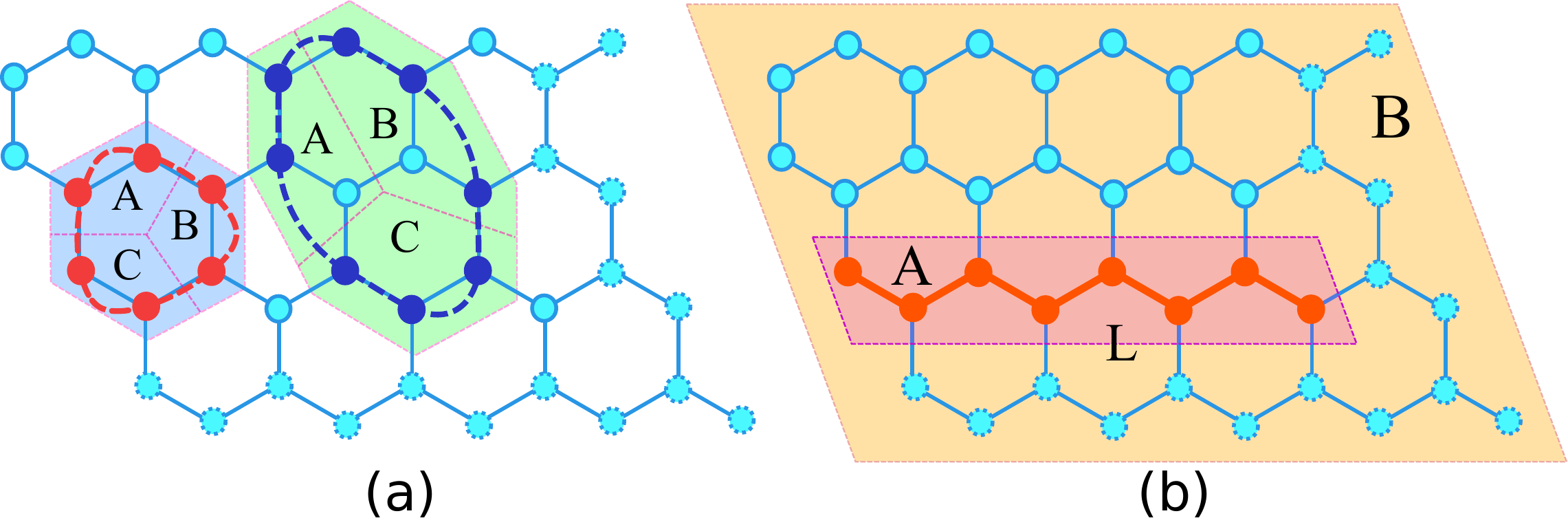}}
\caption{(Color online) (a) Kitaev-Preskill partitioning of the lattice with $N=24$ sites for two different closed strings. The shaded regions are considered as the system for which we calculate the RDM
and the rest of the lattice is regarded as environment. (b) Partitioning of the honeycomb lattice with $N=24$ sites to subsystem A, which is a zigzag chain with length $L$, and the rest of the lattice to subsystem B. }
\label{fig:k_p_partitioning}
\end{figure}
\subsection{Numerical Calculation of TEE}

In this subsection, we calculate the topological entanglement entropy of the TCC numerically by resorting to the Kitaev-Preskill approach, Eq.(\ref{eq:KP-TEE}). 
To this end, we considered different disk shape regions on the honeycomb lattice as our system and the rest of the lattice as environment \cite{Furukawa}
(see blue and green shaded regions in Fig.~\ref{fig:k_p_partitioning}(a)). The entanglement partitions were chosen by ensuring that there exist at least one contractible loop 
inside the system.

Next we divided the disk shape regions to subsystems $A$, $B$ and $C$ according to the KP strategy and calculated the reduced density matrix of 
each subsystem numerically by exact diagonalization (ED) based on the Lanczos method. The numerical diagonalizations were performed 
on honeycomb clusters with $12$, $18$ and $24$ sites and the periodic boundary condition was imposed at the boundaries of the clusters \cite{ssj2}. The RDM of each subsystem
was then extracted from the ground-state wave function of the TCC on the clusters. 
Afterwards, we calculated the TEE of color code by subtracting the von Neumann entropy, $S^{\rm vN}$, of each subsystem according to the KP strategy (see Eq.~\ref{eq:KP-TEE}). Interestingly, our numerical
calculations exactly resulted in $\gamma=2$ for all of the clusters (see yellow diamond in Fig.~\ref{fig:TEE}). This results clearly certify the condensation of
closed strings in the ground state of the system and is a firm proof that the ground state of the TCC is a spin liquid composed of fluctuating loops. 

In order to investigate the area law dependence of the local term of EE, we divided the honeycomb lattice to two subsystems i.e., a zigzag chain with length $L$ as region A and the rest of the lattice as subsystem B (see Fig.~\ref{fig:k_p_partitioning}(b)) and calculated the $S^{\rm vN}$ of the chain by extracting the reduced density matrix of the subsystem A from the 2D wave function of the system on a honeycomb lattice with $24$ sites, for different chain length $L$. The result is shown in the red curve of Fig.~\ref{fig:area_law}. 
Our finding certifies that the entanglement entropy scales linearly ($\alpha=1$)
with the chain size $L$ which is in full correspondence with the analytical results of Eq. (\ref{eq:RenyTCC}). 

Let us stress that such a linear dependence of the local term in
EE to the boundary length holds for disk shape regions either \cite{Furukawa}. However, studying this case was beyond our computational resources and we therefore
limited our calculations to spin chains.

\section{Ground state dependence of TEE}
\label{sec:GRD-TEE}

In previous sections, we outlined that the TEE of an entanglement partition is obtained from the total quantum dimension $\mathcal{D}$ (see Eq.~(\ref{eq:gamma_dim})).
One should note that this statement holds true only if the entanglement partition has a disk shape geometry with contractible boundaries \cite{Vishwanath}. 
However, if the entanglement cut has boundaries with non-trivial loops, such as partitioning the torus into cylinders, the TEE will then depend on the particular linear 
combination of the ground states.

\begin{figure}
\centerline{\includegraphics[width=\columnwidth]{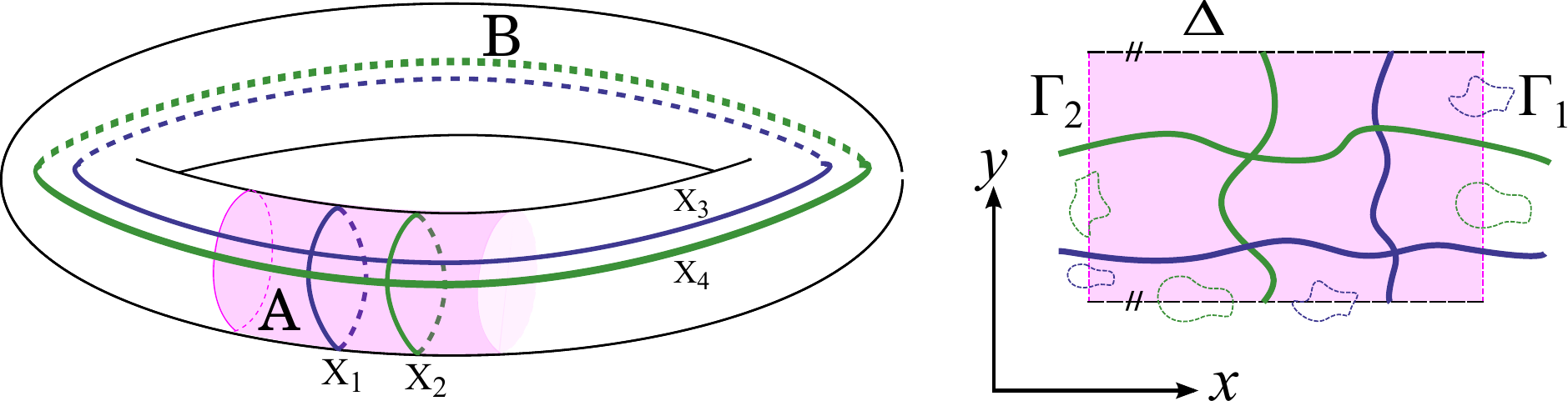}}
\caption{(Color online) Left: bipartitioning the torus into half cylinders. The shaded cylindrical region is considered as subsystem $A$ and the rest is subsystem
$B$. Right: contractible strings intersect the boundaries $\Gamma_1$, $\Gamma_2$ and cut $\Delta$ even number of times, while global loops intersect with them odd number of times. }
\label{fig:non_trivial_cut}
\end{figure}

This can be best understood from the Schmidt decomposition of the ground state of the system on 
a cylindrical bipartition of the torus shown in Fig.~\ref{fig:non_trivial_cut}.
In order to specify the borders and existence of non-trivial loops in the entanglement partition, 
we consider a cut $\Delta$, which goes around the torus in the $x$-direction
and denote the boundaries of the cylindrical subsystem $A$ by $\Gamma_1$, $\Gamma_2$ (see Fig.~\ref{fig:non_trivial_cut}-right).

A direct consequence of the $\Zd \times \Zd$ symmetry of the TCC, which can also be alternatively perceived from Eq.~(\ref{eq:plaquette_superfluous}) is that only two 
out of the three colored strings with colors $c$, $c'$ are independent \cite{bombin_distilation,bombin_two_body}. Making use of this fact, we can fix the following notations for the closed strings crossing the boundaries of the entanglement cut \cite{Vishwanath}.
First, we denote the set of all configurations of independent closed loops lying on the boundaries $\Gamma_1$, $\Gamma_2$ by
$\{\{g^c\},\{g^{c'}\}\}$ where $\{g^c\}\in i=0(1)$ ,$\{g^{c'}\}\in j=0(1)$ if the loop configurations cross the boundaries even (odd) number of times.   
Next, we label the intersection number of the configurations of $c$-strings ($c'$-strings) with the virtual cut $\Delta$ module $2$ by $k=0,1$ ($l=0,1$).
Following this notation, the normalized equal superposition of all the possible configurations of closed loops, $G_{AB}$, in subsystem $A$ ($B$) can be denoted by
$\ket{\psi^{A(B)}_{\{g^c\},\{g^{c'}\},k,l}}$ and therefore a general ground state of the TCC is Schmidt decomposed as
{\footnotesize
\bea
\ket{\xi_{ijkl}}&=& \frac{1}{\sqrt{|G_{AB}|}} \sum_{\substack{ \{\{g^c\},\{g^{c'}\}\}, \\  \{g^c\}\in i,\{g^{c'}\} \in j }} \left(\bigket{\Psi^A_{\{g^c\},\{g^{c'}\},0,0}} \bigket{\Psi^B_{\{g^c\},\{g^{c'}\},k,l}}+\right.  \nonumber \\ 
&& \left. \bigket{\Psi^A_{\{g^c\},\{g^{c'}\},0,1}} \bigket{\Psi^B_{\{g^c\},\{g^{c'}\},k,(l+1){\rm mod}2}} +\right.  \nonumber \\
&& \left. \bigket{\Psi^A_{\{g^c\},\{g^{c'}\},1,0}} \bigket{\Psi^B_{\{g^c\},\{g^{c'}\},(k+1){\rm mod}2,l}} +\right.   \nonumber \\
&& \left. \bigket{\Psi^A_{\{g^c\},\{g^{c'}\},1,1}} \bigket{\Psi^B_{\{g^c\},\{g^{c'}\},(k+1){\rm mod}2,(l+1){\rm mod}2}}\right).
\eea}
As we can clearly see, the key difference with the trivial bipartitioning appears in the Schmidt decomposition of the ground state, 
which originates from the fact that the boundaries of the cylinders are now along the non-contractible loops and we can no longer move them
outside the subregion $A$.

Considering the system to be in a generic state $\ket{\bf \Psi}$ (Eq.~\ref{eq:general_grs}) with an equal weight superposition of the
states from the ground space set, the reduced density matrix of the entanglement cut $A$ is given by
\be
\rho_A={\rm Tr}_B \ket{\bf \Psi}\bra{\bf \Psi},
\ee
where after a very detailed calculation and thereafter diagonalization of the RDM, results in the following Renyi entropy
\bea
S_n &=&\frac{1}{1-n} \log_2 {\rm Tr} (\rho_A^n)  \nonumber \\
&=& \frac{1}{1-n} \log_2 \left((\frac{1}{|G_{AB}|})^{n} |G_{AB}| \sum_{i=i}^{16} p_i^n \right)  \nonumber \\
&=& L \log_2 2 - \gamma'\label{eq:renyi-notrivial},
\eea
where 
\be\label{eq:gammaprime}
\gamma'= 2 \log_2 2 - \frac{1}{1-n} \log_2 \sum_{i=i}^{16} p_i^n
\ee
and $p_i$s are defined as:
\bea
p_1 &=&|a_{0000}+a_{0001}-a_{0010}-a_{0011}|^2 , \nonumber \\
p_2 &=&|a_{0000}-a_{0001}+a_{0010}-a_{0011}|^2 , \nonumber \\
p_3 &=&|a_{0000}-a_{0001}-a_{0010}+a_{0011}|^2 , \nonumber \\
p_4 &=&|a_{0000}+a_{0001}+a_{0010}+a_{0011}|^2 , 
\eea
\bea
p_5 &=&|a_{0100}+a_{0101}-a_{0110}-a_{0111}|^2 , \nonumber \\
p_6 &=&|a_{0100}-a_{0101}+a_{0110}-a_{0111}|^2 , \nonumber \\
p_7 &=&|a_{0100}-a_{0101}-a_{0110}+a_{0111}|^2 , \nonumber \\
p_8 &=&|a_{0100}+a_{0101}+a_{0110}+a_{0111}|^2 , 
\eea
\bea
p_9 &=&|a_{1000}+a_{1001}-a_{1010}-a_{1011}|^2 , \nonumber \\
p_{10} &=&|a_{1000}-a_{1001}+a_{1010}-a_{1011}|^2 , \nonumber \\
p_{11} &=&|a_{1000}-a_{1001}-a_{1010}+a_{1011}|^2 , \nonumber \\
p_{12} &=&|a_{1000}+a_{1001}+a_{1010}+a_{1011}|^2 , 
\eea
\bea
p_{13} &=&|a_{1100}+a_{1101}-a_{1110}-a_{1111}|^2 , \nonumber \\
p_{14} &=&|a_{1100}-a_{1101}+a_{1110}-a_{1111}|^2 , \nonumber \\
p_{15} &=&|a_{1100}-a_{1101}-a_{1110}+a_{1111}|^2 , \nonumber \\
p_{16} &=&|a_{1100}+a_{1101}+a_{1110}+a_{1111}|^2 . 
\eea

As expected, the linear dependence of the local term in EE is evident even for the non-trivial bipartitioning. The difference further
arises in the topological contribution to the entanglement entropy, which is presented here by $\gamma'$. 
The first term in Eq.~(\ref{eq:gammaprime})
is the universal term $\gamma$ originating from the total quantum dimension $\mathcal{D}$ and the second term emerges 
due to the fact that the boundaries of the cylindrical entanglement cut are along the non-contractible loops.

\section{Obtaining minimum entropy states}
\label{sec:MES}
In the previous section, we showed how non-trivial bipartitioning of the torus can affect the entanglement entropy of a topologically ordered system.
Here, we shed light on the concept of minimum entropy states which can later be used to fully characterize a topologically ordered phase.

Given a non-trivial entanglement cut on the torus, the MESs are defined as linear superpositions of the degenerate ground states of the system 
which minimize the entanglement entropy of the given bipartition \cite{Vishwanath,Vidal} i.e. 
\be\label{eq:mes:general}
\ket{\MES^{w}}=\sum_{ijkl} a_{ijkl} \ket{\xi_{ijkl}},
\ee
where $w$ is the direction of the entanglement cut. It can be shown that the MESs are related to the quasiparticles of the model encircling the 
torus perpendicular to the entanglement cut i.e. they are eigenstates of the Wilson loop operators with a certain type of QP \cite{Dong}. 
Later on, we will show how the MESs are related to the quasiparticles and present the procedure of extracting the statistics of QPs from the MESs.

The TCC poses $16$ topologically degenerate ground states and the MESs can therefore be calculated by generating linear superpositions of 
these $16$ states (Eq.~\ref{eq:mes:general}) with minimum entropy. To this end, we first bipartition the torus to two cylinders along the $y$-direction (see Fig.~\ref{fig:non_trivial_cut})
and then calculate the MESs according to the procedure worked out in Ref.~\cite{Vishwanath}. Let us note that minimization of the Renyi EE (Eq.~\ref{eq:renyi-notrivial})
is performed by using the Nelder Mead algorithm \cite{Orus-GME}. The complete set of MESs are further calculated by satisfying the normalization and orthogonalization conditions on
minimum entropy states. Our calculations finally leads to the following MESs for the TCC model: 

\bea
\ket{\MES_1} &=&  \frac{e^{i \phi_1}}{2} (  \ket{\xi_{0000}}-\ket{\xi_{0001}}+\ket{\xi_{0010}}-\ket{\xi_{0011}}),  \\
\ket{\MES_2} &=&  \frac{e^{i \phi_2}}{2} ( -\ket{\xi_{0000}}+\ket{\xi_{0001}}+\ket{\xi_{0010}}-\ket{\xi_{0011}}), \nonumber \\
\ket{\MES_3} &=&  \frac{e^{i \phi_3}}{2} ( -\ket{\xi_{0000}}-\ket{\xi_{0001}}+\ket{\xi_{0010}}+\ket{\xi_{0011}}), \nonumber \\
\ket{\MES_4} &=&  \frac{e^{i \phi_4}}{2} ( -\ket{\xi_{0000}}-\ket{\xi_{0001}}-\ket{\xi_{0010}}-\ket{\xi_{0011}}), \nonumber
\eea
\bea
\ket{\MES_5} &=&  \frac{e^{i \phi_5}}{2} ( \ket{\xi_{0100}}-\ket{\xi_{0101}}+\ket{\xi_{0110}}-\ket{\xi_{0111}}),  \\
\ket{\MES_6} &=&  \frac{e^{i \phi_6}}{2} ( \ket{\xi_{0100}}-\ket{\xi_{0101}}-\ket{\xi_{0110}}+\ket{\xi_{0111}}), \nonumber \\
\ket{\MES_7} &=&  \frac{e^{i \phi_7}}{2} ( -\ket{\xi_{0100}}-\ket{\xi_{0101}}+\ket{\xi_{0110}}+\ket{\xi_{0111}}), \nonumber \\
\ket{\MES_8} &=&  \frac{e^{i \phi_8}}{2} ( -\ket{\xi_{0100}}-\ket{\xi_{0101}}-\ket{\xi_{0110}}-\ket{\xi_{0111}}), \nonumber
\eea
\bea
\ket{\MES_9} &=&  \frac{e^{i \phi_9}}{2} ( \ket{\xi_{1000}}+\ket{\xi_{1001}}+\ket{\xi_{1010}}+\ket{\xi_{1011}}),  \\
\ket{\MES_{10}} &=&  \frac{e^{i \phi_{10}}}{2} ( \ket{\xi_{1000}}-\ket{\xi_{1001}}+\ket{\xi_{1010}}-\ket{\xi_{1011}}), \nonumber \\
\ket{\MES_{11}} &=&  \frac{e^{i \phi_{11}}}{2} ( \ket{\xi_{1000}}-\ket{\xi_{1001}}-\ket{\xi_{1010}}+\ket{\xi_{1011}}), \nonumber \\
\ket{\MES_{12}} &=&  \frac{e^{i \phi_{12}}}{2} ( -\ket{\xi_{1000}}-\ket{\xi_{1001}}+\ket{\xi_{1010}}+\ket{\xi_{1011}}), \nonumber
\eea
\bea
\ket{\MES_{13}} &=&  \frac{e^{i \phi_{13}}}{2} ( \ket{\xi_{1100}}+\ket{\xi_{1101}}+\ket{\xi_{1110}}+\ket{\xi_{1111}}),  \\
\ket{\MES_{14}} &=&  \frac{e^{i \phi_{14}}}{2} ( \ket{\xi_{1100}}-\ket{\xi_{1101}}-\ket{\xi_{1110}}+\ket{\xi_{1111}}), \nonumber \\
\ket{\MES_{15}} &=&  \frac{e^{i \phi_{15}}}{2} ( -\ket{\xi_{1100}}+\ket{\xi_{1101}}-\ket{\xi_{1110}}+\ket{\xi_{1111}}), \nonumber \\
\ket{\MES_{16}} &=&  \frac{e^{i \phi_{16}}}{2} ( -\ket{\xi_{1100}}-\ket{\xi_{1101}}+\ket{\xi_{1110}}+\ket{\xi_{1111}}). \nonumber
\eea
We have further considered an arbitrary phase freedom for each state.

Let us further note that more straightforward approaches for numerical calculations of the MESs by density matrix renormalization group (DMRG) is presented in Ref.~\cite{Balents,sheng}
where they use DMRG to calculate the usual entanglement entropy for the division of a cylinder into two equal halves by a flat cut, and extract the TEE and therefore MESs from its asymptotic, large-circumference limit. A similar strategy based on the infinite projected entangled state (iPEPS) and geometric entanglement (GE) has also worked out in Ref.~\cite{Orus-GME} where the MESs are extracted by minimizing the GE on an infinite cylinder.

In order to perceive the nature of each MES and its relation to the QPs of the model, 
we recall that the QPs are created at the  
end points of the open strings and a particle and its anti-particle are created on the system, 
exciting the corresponding plaquette operators
to their $-1$ eigenvalue. Winding the particle in the $x$- or $y$-direction around the torus and 
bringing it back to its anti-particle
would annihilate the two QPs and bring the system back to its ground state. 
The only difference is that the 
trace of the particle treading is left as a closed loop around the torus, which takes the system to another parity sector defined by that 
global loop. This hold for all charges and fluxes of the $\Zd \times \Zd$ topological phase of the TCC.

We can make use of this fact and define the following Wilson loop operators to detect the charges and fluxes of the corresponding MES \cite{Vishwanath}.
We have already mentioned that two, out of the three colored loops are independent (for example green and blue)
(see Fig.\ref{fig:tcc}). We therefore define $T_y^g$ ($T_x^g$) and $T_y^b$ ($T_x^b$), which detect green and blue magnetic fluxes
by inserting additional electric charge lines in the $y$- ($x$-) direction
\bea
T_x^g\ket{\xi_{0jkl}}&=&\ket{\xi_{1jkl}}, \quad\quad T_x^g\ket{\xi_{1jkl}}=\ket{\xi_{0jkl}}, \\
T_x^b\ket{\xi_{i0kl}}&=&\ket{\xi_{i1kl}}, \quad\quad T_x^b\ket{\xi_{i1kl}}=\ket{\xi_{i0kl}}, 
\eea
\bea
T_y^g\ket{\xi_{ij0l}}&=&\ket{\xi_{ij1l}}, \quad\quad T_y^g\ket{\xi_{ij1l}}=\ket{\xi_{ij0l}}, \\
T_y^b\ket{\xi_{ijk0}}&=&\ket{\xi_{ijk1}}, \quad\quad T_y^b\ket{\xi_{ijk1}}=\ket{\xi_{ijk0}}. 
\eea
The electric charges can further be detected by measuring the phase factor picked up by the charges when winding around the torus perpendicular
to the magnetic field. Similar to the fluxes, there exist two distinct class of charges (blue and green) which are detected by inserting additional magnetic field lines
by the following operators
\bea
F_x^g\ket{\xi_{ijkl}}&=&(-1)^l\ket{\xi_{ijkl}}, \quad F_y^g\ket{\xi_{ijkl}}=(-1)^j\ket{\xi_{ijkl}}, \\
F_x^b\ket{\xi_{ijkl}}&=&(-1)^k\ket{\xi_{ijkl}}, \quad F_y^b\ket{\xi_{ijkl}}=(-1)^i\ket{\xi_{ijkl}}. 
\eea
One can further check that the following algebra holds between the electric and magnetic loop insertion operators:
\bea
T_x^bF_y^g&=&-F_y^g T_x^b, \quad T_x^gF_y^b=-F_y^b T_x^g, \\
T_y^bF_x^g&=&-F_x^g T_y^b, \quad T_y^gF_x^b=-F_x^b T_y^g.
\eea
Applying the loop insertion operators to the MESs of the TCC obtained from a cylindrical cut in the $y$-direction, we arrive at the 
conclusion listed in Table.~\ref{tab:MES_To_QP}.
The MESs are simultaneous eigenstates of the Wilson loop operators $T_y^b$, $T_y^g$, $F_y^b$ and
 $F_y^g$ and their eigenvalues determine the existence of blue and green charges and fluxes on the system as bare QPs.  
Applying the fusion rules (\ref{eq:fusion_rules}) to the cases with more than one bare QPs and clarifying the outcome (or alternatively using Table.~\ref{tab:fusion_table}),
the on-to-one correspondence between the MESs and the 16 quasiparticles (Eq.~\ref{eq:qps}) of the TCC is fully determined.

\begin{table}[!t]
 \begin{center}
 \caption{Correspondence between the MESs of the TCC and the quasiparticles, determined from the action of Wilson loop operators $T_y^b$, $T_y^g$, $F_y^b$ and
 $F_y^g$ on the MESs. }
 \label{tab:MES_To_QP}
   \begin{ruledtabular}
   \begin{tabular}{l|ccccccc}
    MES & $T_y^g$ & $T_y^b$ & $F_y^g$ & $F_y^b$ & Bare QP & Fused QP  \\
    \hline
    $\MES_1$ & 0 & 1 & 0 & 0 & $\chi_m^g$ & $\chi_m^g$ \\
    $\MES_2$ & 1 & 1 & 0 & 0 & $\chi_m^b, \chi_m^g$ & $\chi_m^r$ \\
    $\MES_3$ & 1 & 0 & 0 & 0 & $\chi_m^b$ & $\chi_m^b$ \\    
    $\MES_4$ & 0 & 0 & 0 & 0 & ${\bf 1}$ & ${\bf 1}$ \\    
    $\MES_5$ & 0 & 1 & 0 & 1 & $\chi_m^g, \chi_e^g$ & $\chi_e^g\chi_m^g$ \\    
    $\MES_6$ & 1 & 1 & 0 & 1 & $\chi_m^b, \chi_m^g, \chi_e^g$ & $\chi_e^g\chi_m^r$ \\    
    $\MES_7$ & 1 & 0 & 0 & 1 & $\chi_m^b, \chi_e^g$ & $\chi_e^g\chi_m^b$ \\    
    $\MES_8$ & 0 & 0 & 0 & 1 & $\chi_e^g$ & $\chi_e^b$ \\    
    $\MES_9$ & 0 & 0 & 1 & 0 & $\chi_e^b$ & $\chi_e^g$ \\    
    $\MES_{10}$ & 0 & 1 & 1 & 0 & $\chi_m^g, \chi_e^b$ & $\chi_e^b\chi_m^g$ \\    
    $\MES_{11}$ & 1 & 1 & 1 & 0 & $\chi_m^b, \chi_m^g, \chi_e^b$ & $\chi_e^b\chi_m^r$ \\    
    $\MES_{12}$ & 1 & 0 & 1 & 0 & $\chi_m^b, \chi_e^b$ & $\chi_e^b\chi_m^b$ \\    
    $\MES_{13}$ & 0 & 0 & 1 & 1 & $\chi_e^b, \chi_e^g$ & $\chi_e^r$ \\    
    $\MES_{14}$ & 1 & 1 & 1 & 1 & $\chi_m^b, \chi_m^g, \chi_e^b, \chi_e^g$ & $\chi_e^r\chi_m^r$ \\    
    $\MES_{15}$ & 0 & 1 & 1 & 1 & $\chi_m^g, \chi_e^b, \chi_e^g$ & $\chi_e^r\chi_m^g$ \\    
    $\MES_{16}$ & 1 & 0 & 1 & 1 & $\chi_m^b, \chi_e^b, \chi_e^g$ & $\chi_e^r\chi_m^b$ \\
     \end{tabular}
  \end{ruledtabular}
 \end{center}
\end{table}

\section{Extracting Anyonic braid statistics from MES}
\label{sec:Braid-TEE}
The self and mutual statistics of quasiparticles in an Abelian or non-Abelian topological phase are denoted by $\Um$ and $\Sm$ modular matrices, respectively \cite{Vishwanath,Vidal}.
The elements of the diagonal $\Um$ matrix represent the statistics of a quasiparticle when encircles around itself (alternatively the exchange statistics of two identical QPs)
and the $\Sm_{ij}$ elements of the modular $\Sm$-Matrix denote the mutual statistics of the $i$th quasiparticle with respect to the $j$th one.
Knowing the $\Sm$-matrix, almost a fully characterization of a topologically ordered phase is possible by Verlinde formula \cite{Verlinde} which relates many features of the 
topological phase such as fusion rules and quantum dimension of QPs to the elements of the $\Sm$-Matrix.

The $\Sm$-Matrix further acts as a modular transformation between different MESs defined on entanglement cuts in different directions \cite{Vishwanath}. 
In other word, MESs are the canonical basis for defining the modular matrices i.e.  
\be
\Sm_{ij}=\frac{1}{\mathcal{D}} \braket{\MES^{x}_i}{\MES^{y}_j}.
\ee
When the lattice has $\pi/2$ symmetry, it is shown that 
\be
\Sm_{ij}=\frac{1}{\mathcal{D}} \bra{\MES^{y}_i} R_{\pi/2} \ket{\MES^{y}_j},
\ee
where $R_{\pi/2}$ is the rotation operators acting on the MESs basis. However, for lattices such as honeycomb or Kagome which have $2\pi/3$ symmetry
the above relation is given by
\be\label{eq:Smatrix2pi3}
(D^{\dagger} \Um \Sm D)_{ij}=\frac{1}{\mathcal{D}} \bra{\MES^{y}_i} R_{2\pi/3} \ket{\MES^{y}_j},
\ee
where $D$ is a diagonal matrix with $D_{jj}=e^{i\phi_j}$ corresponding to the arbitrary phase in 
the definition of MESs.

In order to extract the modular matrices of the model,
we need to calculate the action of $R_{2\pi/3}$ on the MESs \cite{Vishwanath}. In order to do so, we first write down the transformation of $\ket{\xi_{ijkl}}$
under $R_{2\pi/3}$ rotation:
\bea
R_{2\pi/3} \ket{\xi_{0000}} &=& \ket{\xi_{0000}},\quad R_{2\pi/3} \ket{\xi_{0001}} = \ket{\xi_{0100}}, \nonumber \\
R_{2\pi/3} \ket{\xi_{0010}} &=& \ket{\xi_{1000}},\quad R_{2\pi/3} \ket{\xi_{0011}} = \ket{\xi_{1100}}, \nonumber \\
R_{2\pi/3} \ket{\xi_{0100}} &=& \ket{\xi_{0101}},\quad R_{2\pi/3} \ket{\xi_{0101}} = \ket{\xi_{0001}}, \nonumber \\
R_{2\pi/3} \ket{\xi_{0110}} &=& \ket{\xi_{1101}},\quad R_{2\pi/3} \ket{\xi_{0111}} = \ket{\xi_{1001}}, \nonumber \\
R_{2\pi/3} \ket{\xi_{1000}} &=& \ket{\xi_{1010}},\quad R_{2\pi/3} \ket{\xi_{1001}} = \ket{\xi_{1110}}, \nonumber \\
R_{2\pi/3} \ket{\xi_{1010}} &=& \ket{\xi_{0010}},\quad R_{2\pi/3} \ket{\xi_{1011}} = \ket{\xi_{0110}}, \nonumber \\
R_{2\pi/3} \ket{\xi_{1100}} &=& \ket{\xi_{1111}},\quad R_{2\pi/3} \ket{\xi_{1101}} = \ket{\xi_{1011}}, \nonumber \\
R_{2\pi/3} \ket{\xi_{1110}} &=& \ket{\xi_{0111}},\quad R_{2\pi/3} \ket{\xi_{1111}} = \ket{\xi_{0011}}. 
\eea

Translating the action of  $R_{2\pi/3}$ on MESs, product of the modular matrices, Eq.~(\ref{eq:Smatrix2pi3}), of the TCC on the honeycomb lattice is obtained.
Following the procedure described in appendix \ref{AP:modular_matrices}, the $\Um$ and $\Sm$ modular matrices of the TCC are extracted in a straightforward manner 
(see Eq.~(\ref{eq:S_U_Matrix}) for the explicit form of the $\Um$ and $\Sm$ matrices).

As we have already mentioned, the elements of the $\Um$-Matrix represent the self statistics of the QPs. As we can see, $6$ out of the $16$ diagonal 
elements of the $\Um$-Matrix are $-1$, which denote the fermions and the rest of the elements represent the bosons with $+1$ self statistics. 
The elements of the $\Sm$-Matrix further represent the mutual statistics of the QPs. 
One should note that there is a degree of freedom for the location of the elements of 
the $\Um$ and $\Sm$ matrices, which depends on the arrangement of the MESs and the order we 
choose to build the product matrix (\ref{eq:Smatrix2pi3}). Alternatively, one can use the Verlinde \cite{Verlinde} formula to extract the elements of the $\Sm$-Matrix and
quantum dimension of the underlying Abelian theory as well as the fusion rules of the TCC model.

\section{Robustness of topological order}
\label{sec:QPT-TEE}

\begin{figure}[t!]
\centerline{\includegraphics[width=\columnwidth, height=6cm]{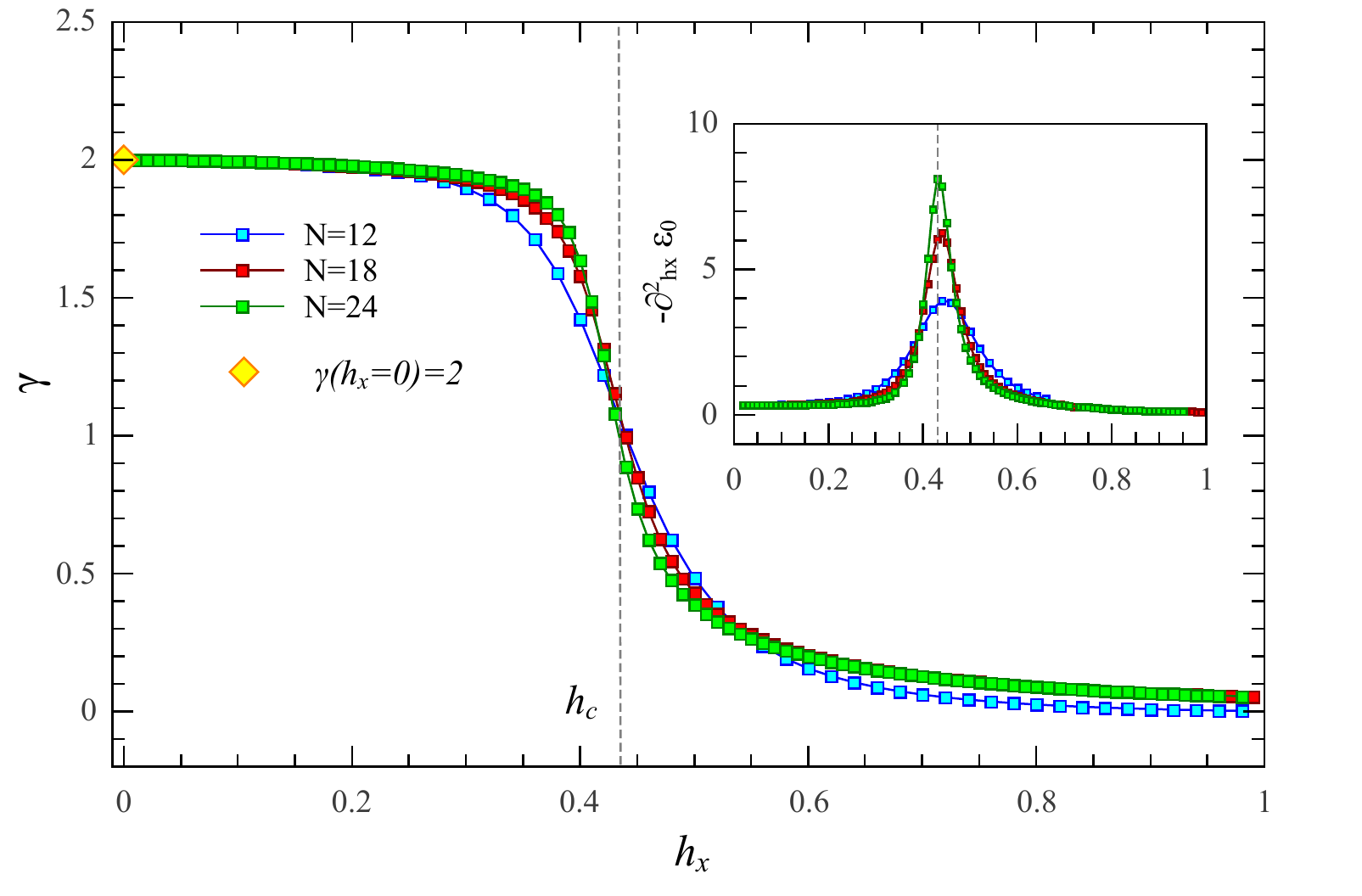}}
\caption{(Color online) Topological entanglement entropy of the TCC as a function of magnetic field $h_x$ for a disk shape region similar to the blue region in 
Fig.~\ref{fig:k_p_partitioning} for clusters with different sites. The inset demonstrates the second derivative of the ground state energy per site.
} 
\label{fig:TEE}
\end{figure}

Topologically ordered phases are renowned for their robust ground state against local perturbations, which can be  used as a reliable resource for storing quantum information
and quantum computation. Robustness of the topologically ordered phases in the presence of external perturbations has been the subject of several 
research papers in the past years, which mostly resort to the study of
energy spectrum of the system to capture the possible phase transitions and breakdown of topological order.
For example, it has been shown that the overall phase diagram of the toric code in the presence of magnetic field is very rich containing first- and second-order phase transitions,
multicriticality, and self-duality depending on the field direction. If the transition is second order, it is typically in the 3D Ising universality class except 
on a special line in parameter space where a more complicated behavior is detected \cite{toric_terbest,toric_hamma,toric_vidal1,toric_vidal2,toric_tupitsyn,toric_dusuel,toric_wu,toric_kps,toric_schulz}.

The TCC in the presence of magnetic field with arbitrary direction undergoes first-order phase transition for all field directions. In contrast, if instead of magnetic filed, 
we choose the Ising interactions with ($j_x,j_y,j_z$) couplings as perturbation, we capture
a second-order quantum phase transition to a $\Zd$ symmetry-broken phase for Ising interactions ($j_x,j_z$), while a first-order transition is
found for a pure interaction $j_y$. The universality is typically 3D
Ising. However, a different critical behavior is found on a multicritical line with $j_x=j_z$ and finite $j_y$ where
critical exponents appear to vary continuously which is very similar to the behavior found for the toric code in a field.
Interestingly, our results for this isotropic plane ($j_x=j_z$) suggest the existence of a first-order plane and a gapless phase
which is adiabatically linked to the gapless $U(1)$ symmetry broken $XY$ model in the limit of large Ising interactions \cite{ssj,ssj2}.

One should note that many intersecting features of 
TO is hidden in the ground state of system and solely observing the energy spectrum 
would not suffice to gain a clear insight about the breakdown of a topological phase.
A clever idea is to utilize quantities, which are directly calculated from ground 
state wave function to characterize a topological phase and possible phase transitions. 
Topological entanglement entropy, entanglement spectrum and ground-state fidelity are several examples
of such measures to distinguish between a topological phase and conventional phases.

In this section, we study the robustness of the TCC in the presence of a parallel magnetic field by calculating the TEE, ES and fidelity.
The Hamiltonian of the color code in the presence of a parallel magnetic field is given by \cite{ssj}:
\be\label{eq:HTCC-Field}
H=-J\sum_{p \in \Lambda} (X_p+Z_p)-h_x \sum_{i} \sigma_i^x,
\ee
where the first term is $H_{\rm TCC}$ and the second term is a uniform magnetic field 
in the $x$-direction, which acts on every vertex $i$ of the lattice $\Lambda$.
Without loss of generality, here we set $h_x>0$.
The parallel field dos not commute with the $Z_p$ plaquette operators and 
consequently the Hamiltonian (\ref{eq:HTCC-Field}) is no longer exactly solvable.

In the extreme limit where the magnetic field is absent ($h_x=0$) the ground state of the system is a spin liquid with $\Zd \times \Zd$ topological order. However,
for $J=0$, the system is in a polarized phase pointing in the $x$-direction. 
It is therefore reasonable to expect a quantum phase transition between these two phases.

The $X_p$ plaquette operators commute with the field term and the full Hamiltonian (\ref{eq:HTCC-Field}) and the eigenvalues $x_p=\pm1$ are conserved quantities
in the presence of the magnetic field. In Ref. \cite{ssj}, it was shown that 
the TCC in the parallel magnetic field is mapped to the Baxter-Wu model in a 
transverse field and the breakdown of the TO in color code was studied by investigating 
the spectral properties of the mapped model. However, one should note that the Baxter-Wu 
model is not topologically ordered \cite{ssj3} and although probing the energetics of 
the model provide us with useful information about the quantum phase transition in the 
TCC in the parallel field, it fails to answer the important question that what doses 
happen to the TO and the ground-state wave function when tuning the magnetic field?

\begin{figure}[t!]
\centerline{\includegraphics[width=\columnwidth,  height=6cm]{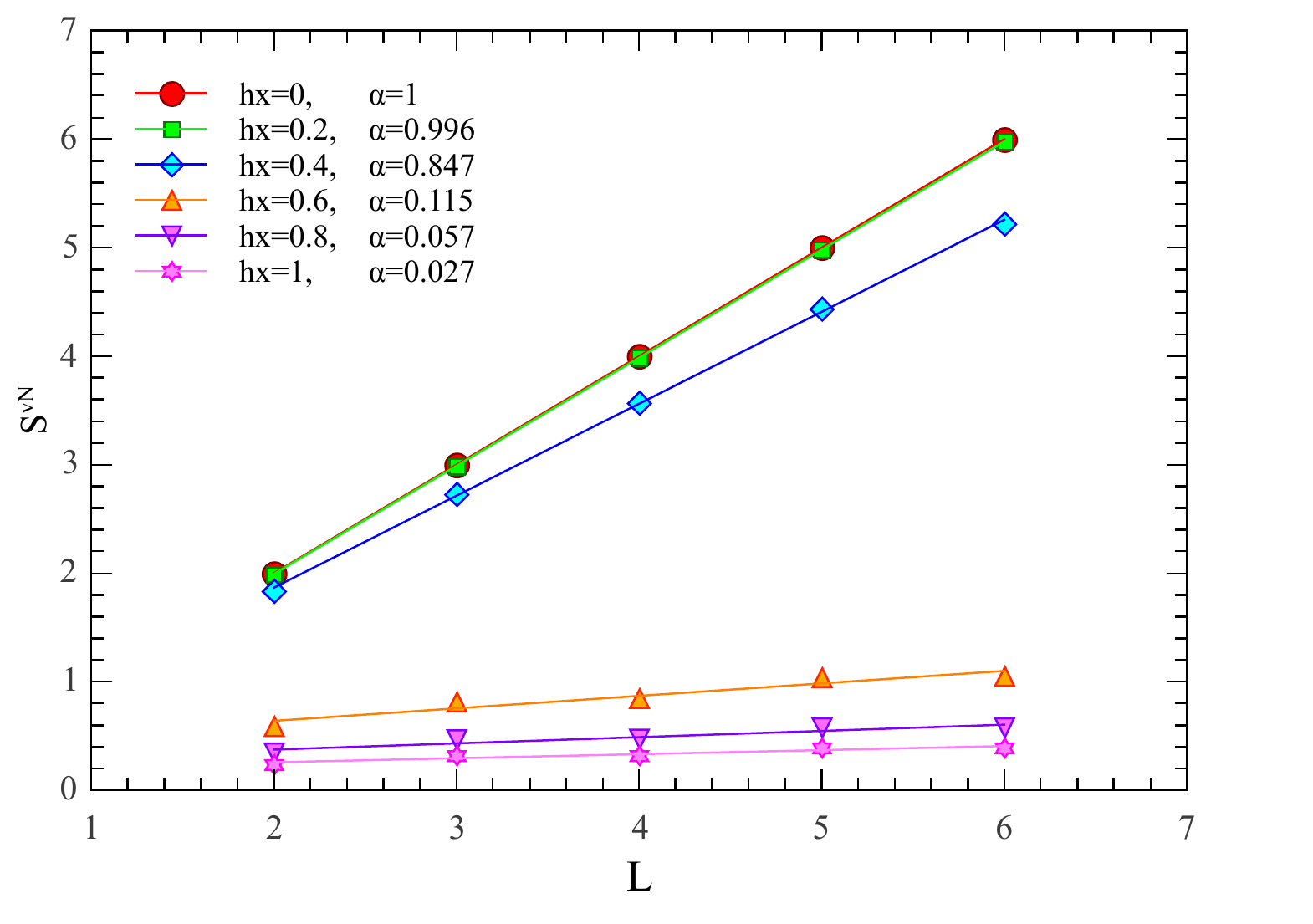}}
\caption{(Color online) Linear dependence of the entanglement entropy to the boundary length $L$ for different values of magnetic field for a disk shape region 
similar to the blue region in Fig.~\ref{fig:k_p_partitioning} for clusters with $24$ sites. }
\label{fig:area_law}
\end{figure}

In order to respond to this question, we have calculated the TEE of color code as function of the magnetic field shown in Fig.~\ref{fig:TEE}.
As we have already outlined, the nonzero $\gamma$ is a clear signature of the TO originating from the presence of loop structures in the ground state. Fig.~\ref{fig:TEE}
shows that the TEE starts from $2$ for the pure TCC at $h_x=0$ and continues to stay close to this value until $h_c\approx 0.43$ which suddenly drops to zero, signaling 
a clear transition from a topologically ordered phase to a polarized phase. Fig.~\ref{fig:TEE} further certifies that the topologically ordered ground state of the TCC is robust against magnetic perturbations and the TO persists until the vicinity of the
transition point where the strength of the magnetic field finally overcome the string's tension and align the spins in the field direction.
Additionally, our calculation on clusters with different size reveals that 
in the thermodynamic limit ($N\rightarrow \infty$)
the TEE behaves like a step function, which jumps from $2$ to $0$ at the
critical point.
However, the increment of correlation length close to the critical point leads to a continuous
change of $\gamma$ with a steep slope for a finite size cluster.
Let us note that the transition point determined from the TEE is in full agreement 
with the $h_c$ obtained from the second derivative of the ground state energy (see the
inset of Fig.~\ref{fig:TEE}). 


We have further studied the boundary length dependence of the entanglement entropy, 
$S^{\rm vN}$, for a spin chain being a subsystem as function of the magnetic field 
in Fig.~\ref{fig:area_law}. 
The results show that the local term of the EE scales linearly with boundary length, $L$, 
even in the presence of the magnetic field as far as we are 
in the topological phase, $h_x<h_c$. Here, the $\alpha$ 
value stays close to $1$ and deflects slightly from each other, signaling that the system is in a robust topological phase. However, on the other side of the transition point  $h_x>h_c$
the $\alpha$ parameter suddenly decrease dramatically to values close to zero signaling a non-entangled polarized phase. This can be further perceived visually from 
Fig.~\ref{fig:area_law}, i.e. the sharp distant between the three upper plots and the lower plots in the figure clearly pinpoints that the plots belongs to two different phases. 

\begin{figure}[t!]
\centerline{\includegraphics[width=\columnwidth,  height=6cm]{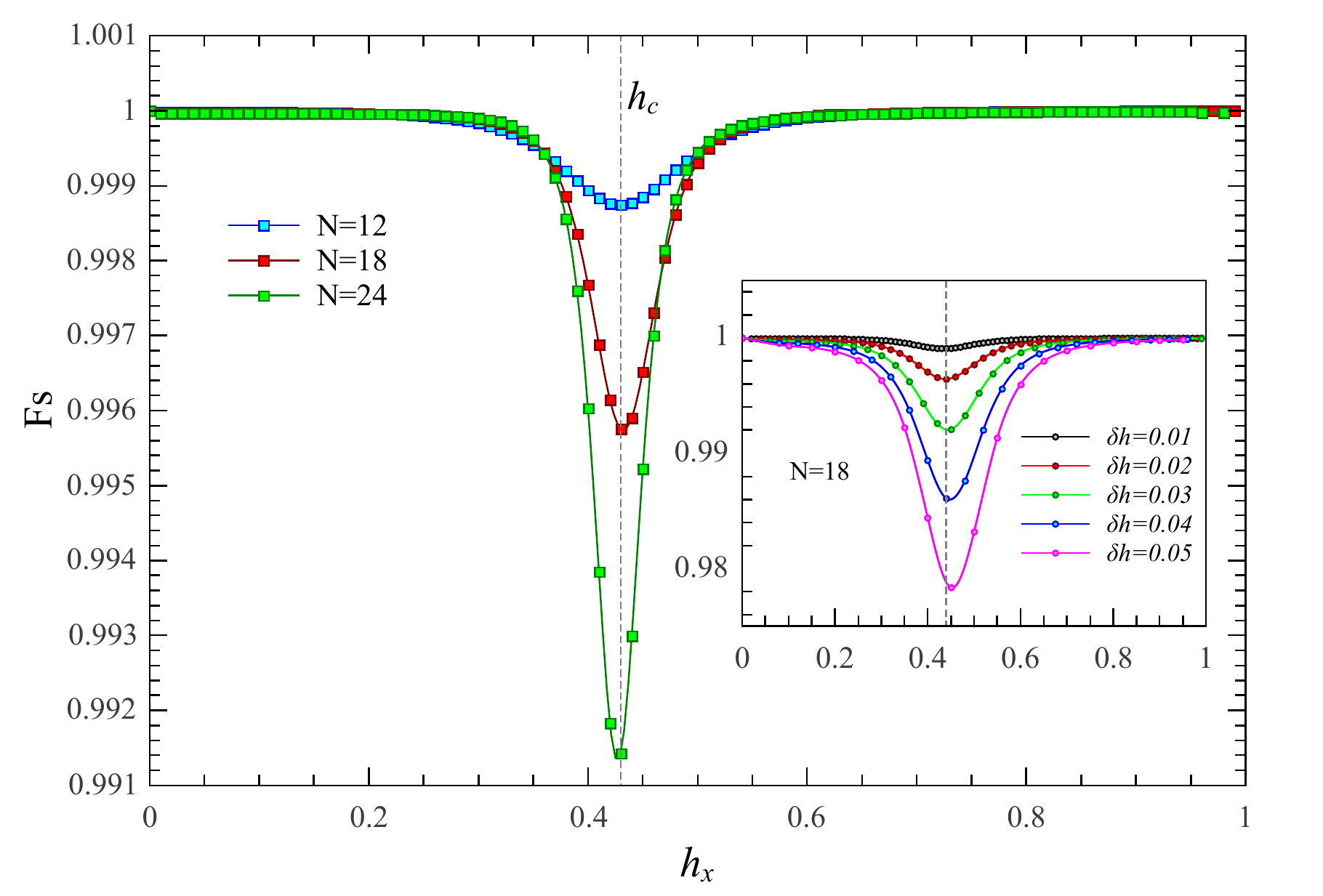}}
\caption{(Color online) Ground state fidelity of the TCC in the presence of the parallel magnetic field for 
$\delta h_x=0.01$ on clusters with different sites.  The sudden drop of the fidelity 
signals the phase transition. The inset depicts fidelity for different $\delta h_x=0.01$, which 
shows the distance between two states is increased by increasing $\delta h_x$.
}
\label{fig:fidelity}
\end{figure}

As another probe to monitor the ground state wave-function, we have calculated the ground state fidelity. 
Considering a Hamiltonian of the form $H(\lambda)=H_0+\lambda H_1$
where $[H_0, H_1]\neq 0$  and $\lambda$ is a coupling to represent the strength of $H_1$.
Hence, the ground state of $H(\lambda)$ depends on the value of $\lambda$, 
i.e.$H(\lambda) |\psi_0\rangle = E_0 |\psi_0\rangle$.
The ground-state fidelity is a measure of the change in the ground state as a result of a 
slight change ($\delta \lambda$)
in the parameter $\lambda$, which is defined by
(\cite{fidelity} and references therein)
\be
Fs=|\braket{\psi_0(\lambda)}{\psi_0(\lambda+\delta \lambda)}|.
\ee
At a quantum critical point, the nature of ground state would change drastically, which
leads to a drop in the ground-state fidelity.
Fig.~\ref{fig:fidelity} demonstrates the ground state fidelity of the TCC in the presence of the parallel magnetic field for $\delta h_x=0.01$ on honeycomb 
clusters with different sizes calculated by ED. 
Evidently, the quantum critical point is marked by a sudden drop of fidelity. This behavior can be ascribed to a dramatic change in the 
structure of the ground state of the system from a topologically ordered phase to a polarized one. 

The fidelity is a measure of the angle distant between two states \cite{fidelity} i.e. the larger the magnitude of the fidelity is, the more distant the two states would be. 
In the vicinity of the critical point, $F_s$ starts to decrease implying that the states are becoming more distant and the nature of the ground state is changing as a 
function magnetic field. The left inset of Fig.~\ref{fig:fidelity} depicts the variation of $F_s$ for 
different $\delta h_x$ for cluster with $N=18$, which provides
by increasing $\delta h_x$, the distant between the two ground states is increased either and $F_s$ drops sharper.   
In the thermodynamic limit, the fidelity might be zero which implies that the two states are totally 
orthogonal no matter how small the $\delta h_x$ parameter is, i.e. the ortogonality catastrophe.

%

\begin{figure}[t!]
\centerline{\includegraphics[width=\columnwidth,  height=6cm]{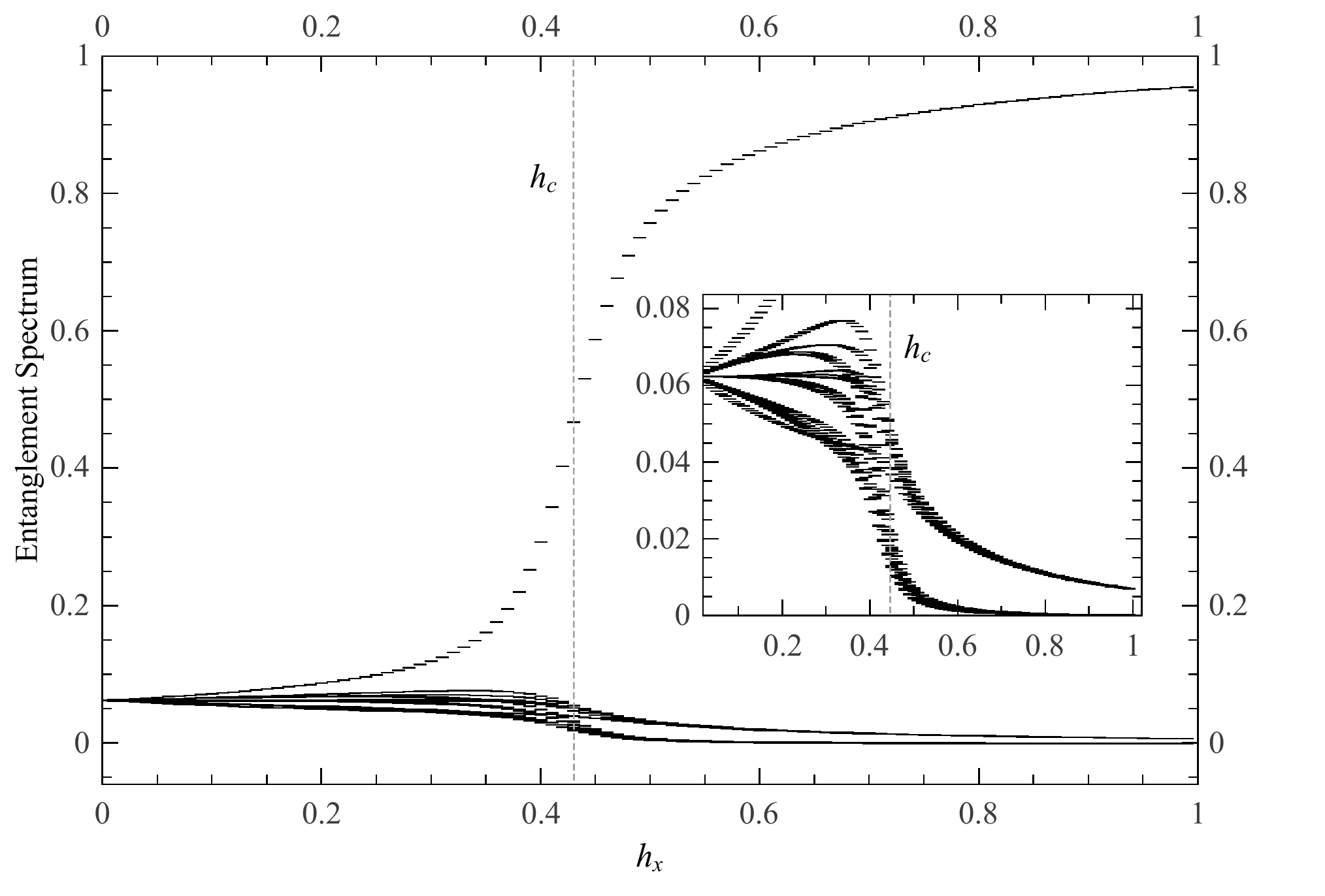}}
\caption{(Color online) Entanglement spectrum of the TCC as a function of magnetic field $h_x$ for a disk shape region similar to the blue region in 
Fig.~\ref{fig:k_p_partitioning} for clusters with $24$ sites. The ES starts to split in the vicinity of the transition point $h_c\approx 0.43$.}
\label{fig:spectrum}
\end{figure}

The quantum phase transition of the $\Zd \times \Zd$ topological phase of TCC to the polarized phase can further 
be captured by looking at the entanglement spectrum
of the system (see Fig.~\ref{fig:spectrum}). One can clearly see that the spectrum of the RDM of a disk shape region 
(blue one in Fig.~\ref{fig:k_p_partitioning})
splits in the proximity of the transition point, which is a result of quantum fluctuations in 
all length scales, i.e. a critical behavior. The change of RDM pattern as shown in Fig.~\ref{fig:spectrum}
can be thought as another signal 
witnessing the phase transition at $h_c\approx 0.43$.

\section{Summary and Conclusions}
\label{sec:conclude}

We studied the entanglement entropy of the topological color code on the honeycomb lattice wrapped around a torus of genus $g=1$, 
both analytically and numerically. Our analytical approach relied on calculating the Renyi entanglement entropy for a disk 
shape regions with contractible boundary. We found that the EE of the TCC has 
a local contribution which scales linearly with boundary of the entanglement partition and further 
has a topological contribution i.e. the topological entanglement entropy
which is universal and is related to the total quantum dimension of the underlying Abelian theory, 
which for the TCC with $\mathcal{D}=4$ is obtained to be $\gamma=\log_2 \mathcal{D}=2$.

Our Numerical approach was based on finite size exact diagonalization on periodic honeycomb clusters 
with $12$, $18$ and $24$ sites by which we calculated the von Neumann entropy of the 
ground state, and implementing Kitaev-Preskill strategy to calculate TEE. 
Interestingly, our findings are in full agreement with analytical results. 

Furthermore, we investigated the ground state dependence of the TEE by calculating the Renyi entropy on regions with non-contractible 
boundaries, i.e. by partitioning the torus to cylindrical 
subregions and showed that the ground state wave function of the 
system is Schmidt decomposed differently on such a partitioning and the TEE will therefore depend on 
the chosen ground state. Aside from that, we identified
the minimum entropy states of the TCC for the non-trivial partitioning of the torus and related the 
MESs to the corresponding QPs, by defining certain types of Wilson loop operators.
We eventually,  extracted the $\Um$ and $\Sm$ modular matrices of the TCC from MESs and determined
the self and mutual braid statistics of the anyonic quasiparticles of the system and fully characterized 
the $\Zd \times \Zd$ Abelian phase of the TCC model.

Aside from that, the minimum entropy states, provided in the paper, are highly invaluable not only for determining the modular matrices
of the TCC, but also for studying the concept of symmetry enriched topological (SET) phases \cite{SET1,SET2,SET3,SET4}. Our results open the door for answering to this 
important question as to whether the TCC is a SET phase or not and what kind of symmetry is responsible for the symmetry fractionalization of the 
topological order in color codes and emergence of fractional quasiparticles in the model.

Due to the lack of local order parameter in topological phases, 
we resorted to the measures obtained from ground state wave function, to probe the stability of the 
topological order of TCC in the presence of a parallel magnetic field in the $x$-direction. 
We therefore calculated the TEE, entanglement spectrum and ground state fidelity
of the TCC as a function of magnetic field from the numerically obtained ground state on honeycomb clusters.
Our findings reveal that the TEE, which is a clear signature of TO, starts from $2$ for the pure TCC at $h_x=0$ 
and continues to stay close to this value up until $h_c\approx 0.43$, which suddenly drops to
zero, signaling a clear transition from a topologically ordered phase to a trivial (polarized) one. 
Moreover, we find that the topologically ordered ground state is 
robust against magnetic perturbations and the TO persists until the vicinity of the transition point, 
where the strength of the magnetic field finally overcome the string's 
tension and align the spins in the field direction. The quantum critical point detected from monitoring 
the ground state wave function is in full agreement with the previous results \cite{ssj2} 
acquired from analysis of the energy spectrum.

We further found that the ground state fidelity, which is a measure of angle distant between two states, 
stays close to $1$ and sharply drops in the proximity of the quantum phase transition
revealing the robust nature of the TO in TCC in the presence of magnetic field, 
as well as the location of the critical point, which is in full agreement with TEE.

Eventually, we found that the entanglement spectrum of the RDM of the TCC splits severely close to the 
transition point, which is another proof for the change in the 
nature of the phases in system, while tuning the magnetic field.

\section{Acknowledgements}
This work is supported by Iran National Science Foundation (INSF) under Grant NO. 93023859 and 
partly by Sharif University of Technology's Office of Vice President for Research.
 
\onecolumngrid
\appendix

\section{Fusion table of the quasiparticle excitations of TCC}
\label{AP:fusion}
Using the fusion rules provided in Eq.~(\ref{eq:fusion_rules}), the fusion table of the quasiparticle excitationas of the color code is given by
\begin{center}
\begin{table}[!htbp]
\setcounter{table}{0}
\renewcommand\thetable{\thesection.\arabic{table}} 
\caption{Fusion table of the topological color code}
\begin{tabular}{l|l*{16}{c}r}
$\bigotimes$ & ${\bf 1}$ & $\chi_e^r$ & $\chi_e^g$ & $\chi_e^b$ & $\chi_m^r$ & $\chi_m^g$ & $\chi_m^b$ & $\chi_{e}^{r} \chi_{m}^{r}$ & $\chi_{e}^{g} \chi_{m}^{g}$ & $\chi_{e}^{b} \chi_{m}^{b}$ & $\chi_{e}^{r} \chi_{m}^{g}$ & $\chi_{e}^{g} \chi_{m}^{r}$ & $\chi_{e}^{r} \chi_{m}^{b}$ & $\chi_{e}^{b} \chi_{m}^{r}$ & $\chi_{e}^{b} \chi_{m}^{g}$ & $\chi_{e}^{g} \chi_{m}^{b}$ \\
\hline
${\bf 1}$ &  ${\bf 1}$ & $\chi_e^r$ & $\chi_e^g$ & $\chi_e^b$ & $\chi_m^r$ & $\chi_m^g$ & $\chi_m^b$ & $\chi_{e}^{r} \chi_{m}^{r}$ & $\chi_{e}^{g} \chi_{m}^{g}$ & $\chi_{e}^{b} \chi_{m}^{b}$ & $\chi_{e}^{r} \chi_{m}^{g}$ & $\chi_{e}^{g} \chi_{m}^{r}$ & $\chi_{e}^{r} \chi_{m}^{b}$ & $\chi_{e}^{b} \chi_{m}^{r}$ & $\chi_{e}^{b} \chi_{m}^{g}$ & $\chi_{e}^{g} \chi_{m}^{b}$ \\
$\chi_e^r$ &  $\chi_e^r$ & ${\bf 1}$ & $\chi_e^b$ & $\chi_e^g$ & $\chi_{e}^{r} \chi_{m}^{r}$ & $\chi_{e}^{r} \chi_{m}^{g}$ & $\chi_{e}^{r} \chi_{m}^{b}$ & $\chi_m^r$ & $\chi_{e}^{b} \chi_{m}^{g}$ & $\chi_{e}^{g} \chi_{m}^{b}$ & $\chi_m^g$ & $\chi_{e}^{b} \chi_{m}^{r}$ & $\chi_m^b$ & $\chi_{e}^{g} \chi_{m}^{r}$ & $\chi_{e}^{g} \chi_{m}^{g}$ & $\chi_{e}^{b} \chi_{m}^{b}$ \\
$\chi_e^g$ &  $\chi_e^g$ & $\chi_e^b$ & ${\bf 1}$ & $\chi_e^r$ & $\chi_{e}^{g} \chi_{m}^{r}$ & $\chi_{e}^{g} \chi_{m}^{g}$ & $\chi_{e}^{g} \chi_{m}^{b}$ & $\chi_{e}^{b} \chi_{m}^{r}$ & $\chi_m^g$ & $\chi_{e}^{r} \chi_{m}^{b}$ & $\chi_{e}^{b} \chi_{m}^{g}$ & $\chi_m^r$ & $\chi_{e}^{b} \chi_{m}^{b}$ & $\chi_{e}^{r} \chi_{m}^{r}$ & $\chi_{e}^{r} \chi_{m}^{g}$ & $\chi_m^b$ \\
$\chi_e^b$ &  $\chi_e^b$ & $\chi_e^g$ & $\chi_e^r$ & ${\bf 1}$ & $\chi_{e}^{b} \chi_{m}^{r}$ & $\chi_{e}^{b} \chi_{m}^{g}$ & $\chi_{e}^{b} \chi_{m}^{b}$ & $\chi_{e}^{g} \chi_{m}^{r}$ & $\chi_{e}^{r} \chi_{m}^{g}$ & $\chi_m^b$ & $\chi_{e}^{g} \chi_{m}^{g}$ & $\chi_{e}^{r} \chi_{m}^{r}$ & $\chi_{e}^{g} \chi_{m}^{b}$ & $\chi_m^r$ & $\chi_m^g$ & $\chi_{e}^{r} \chi_{m}^{b}$ \\
$\chi_m^r$ &  $\chi_m^r$ & $\chi_{m}^{r} \chi_{e}^{r}$ & $\chi_{m}^{r} \chi_{e}^{g}$ & $\chi_{m}^{r} \chi_{e}^{b}$ & ${\bf 1}$ & $\chi_m^b$ & $\chi_m^g$ & $\chi_e^r$ & $\chi_{e}^{g} \chi_{m}^{b}$ & $\chi_{e}^{b} \chi_{m}^{g}$ & $\chi_{e}^{r} \chi_{m}^{b}$ & $\chi_e^g$ & $\chi_{e}^{r} \chi_{m}^{g}$ & $\chi_e^b$ & $\chi_{e}^{b} \chi_{m}^{b}$ & $\chi_{e}^{g} \chi_{m}^{g}$ \\
$\chi_m^g$ &  $\chi_m^g$ & $\chi_{m}^{g} \chi_{e}^{r}$ & $\chi_{m}^{g} \chi_{e}^{g}$ & $\chi_{m}^{g} \chi_{e}^{b}$ & $\chi_m^b$ & ${\bf 1}$ & $\chi_m^r$ & $\chi_{e}^{r} \chi_{m}^{b}$ & $\chi_e^g$ & $\chi_{e}^{b} \chi_{m}^{r}$ & $\chi_e^r$ & $\chi_{e}^{g} \chi_{m}^{b}$ & $\chi_{e}^{r} \chi_{m}^{r}$ & $\chi_{e}^{b} \chi_{m}^{b}$ & $\chi_e^b$ & $\chi_{e}^{g} \chi_{m}^{r}$ \\
$\chi_m^b$ &  $\chi_m^b$ & $\chi_{m}^{b} \chi_{e}^{r}$ & $\chi_{m}^{b} \chi_{e}^{g}$ & $\chi_{m}^{b} \chi_{e}^{b}$ & $\chi_m^g$ & $\chi_m^r$ & ${\bf 1}$ & $\chi_{e}^{r} \chi_{m}^{g}$ & $\chi_{e}^{g} \chi_{m}^{r}$ & $\chi_e^b$ & $\chi_{e}^{r} \chi_{m}^{r}$ & $\chi_{e}^{g} \chi_{m}^{g}$ & $\chi_e^r$ & $\chi_{e}^{b} \chi_{m}^{g}$ & $\chi_{e}^{b} \chi_{m}^{r}$ & $\chi_e^g$ \\
$\chi_{e}^{r} \chi_{m}^{r}$ &  $\chi_{e}^{r} \chi_{m}^{r}$ & $\chi_m^r$ & $\chi_{e}^{b} \chi_{m}^{r}$ & $\chi_{e}^{g} \chi_{m}^{r}$ & $\chi_e^r$ & $\chi_{e}^{r} \chi_{m}^{b}$ & $\chi_{e}^{r} \chi_{m}^{g}$ & ${\bf 1}$ & $\chi_{e}^{b} \chi_{m}^{b}$ & $\chi_{e}^{g} \chi_{m}^{g}$ & $\chi_m^b$ & $\chi_e^b$ & $\chi_m^g$ & $\chi_e^g$ & $\chi_{e}^{g} \chi_{m}^{b}$ & $\chi_{e}^{b} \chi_{m}^{g}$ \\
$\chi_{e}^{g} \chi_{m}^{g}$ &  $\chi_{e}^{g} \chi_{m}^{g}$ & $\chi_{e}^{b} \chi_{m}^{g}$ & $\chi_m^g$ & $\chi_{e}^{r} \chi_{m}^{g}$ & $\chi_{e}^{g} \chi_{m}^{b}$ & $\chi_e^g$ & $\chi_{e}^{g} \chi_{m}^{r}$ & $\chi_{e}^{b} \chi_{m}^{b}$ & ${\bf 1}$ & $\chi_{e}^{r} \chi_{m}^{r}$ & $\chi_e^b$ & $\chi_m^b$ & $\chi_{e}^{b} \chi_{m}^{r}$ & $\chi_{e}^{r} \chi_{m}^{b}$ & $\chi_e^r$ & $\chi_m^r$ \\
$\chi_{e}^{b} \chi_{m}^{b}$ &  $\chi_{e}^{b} \chi_{m}^{b}$ & $\chi_{e}^{g} \chi_{m}^{b}$ & $\chi_{e}^{r} \chi_{m}^{b}$ & $\chi_m^b$ & $\chi_{e}^{b} \chi_{m}^{g}$ & $\chi_{e}^{b} \chi_{m}^{r}$ & $\chi_e^b$ & $\chi_{e}^{g} \chi_{m}^{g}$ & $\chi_{e}^{r} \chi_{m}^{r}$ & ${\bf 1}$ & $\chi_{e}^{g} \chi_{m}^{r}$ & $\chi_{e}^{r} \chi_{m}^{g}$ & $\chi_e^g$ & $\chi_m^g$ & $\chi_m^r$ & $\chi_e^r$ \\
$\chi_{e}^{r} \chi_{m}^{g}$ &  $\chi_{e}^{r} \chi_{m}^{g}$ & $\chi_m^g$ & $\chi_{e}^{b} \chi_{m}^{g}$ & $\chi_{e}^{g} \chi_{m}^{g}$ & $\chi_{e}^{r} \chi_{m}^{b}$ & $\chi_e^r$ & $\chi_{e}^{r} \chi_{m}^{r}$ & $\chi_m^b$ & $\chi_e^b$ & $\chi_{e}^{g} \chi_{m}^{r}$ & ${\bf 1}$ & $\chi_{e}^{b} \chi_{m}^{b}$ & $\chi_m^r$ & $\chi_{e}^{g} \chi_{m}^{b}$ & $\chi_e^g$ & $\chi_{e}^{b} \chi_{m}^{r}$ \\
$\chi_{e}^{g} \chi_{m}^{r}$ &  $\chi_{e}^{g} \chi_{m}^{r}$ & $\chi_{e}^{b} \chi_{m}^{r}$ & $\chi_m^r$ & $\chi_{e}^{r} \chi_{m}^{r}$ & $\chi_e^g$ & $\chi_{e}^{g} \chi_{m}^{b}$ & $\chi_{e}^{g} \chi_{m}^{g}$ & $\chi_e^b$ & $\chi_m^b$ & $\chi_{e}^{r} \chi_{m}^{g}$ & $\chi_{e}^{b} \chi_{m}^{b}$ & ${\bf 1}$ & $\chi_{e}^{b} \chi_{m}^{g}$ & $\chi_e^r$ & $\chi_{e}^{r} \chi_{m}^{b}$ & $\chi_m^g$ \\
$\chi_{e}^{r} \chi_{m}^{b}$ &  $\chi_{e}^{r} \chi_{m}^{b}$ & $\chi_m^b$ & $\chi_{e}^{b} \chi_{m}^{b}$ & $\chi_{e}^{g} \chi_{m}^{b}$ & $\chi_{e}^{r} \chi_{m}^{g}$ & $\chi_{e}^{r} \chi_{m}^{r}$ & $\chi_e^r$ & $\chi_m^g$ & $\chi_{e}^{b} \chi_{m}^{r}$ & $\chi_e^g$ & $\chi_m^r$ & $\chi_{e}^{b} \chi_{m}^{g}$ & ${\bf 1}$ & $\chi_{e}^{g} \chi_{m}^{g}$ & $\chi_{e}^{g} \chi_{m}^{r}$ & $\chi_e^b$ \\
$\chi_{e}^{b} \chi_{m}^{r}$ &  $\chi_{e}^{b} \chi_{m}^{r}$ & $\chi_{e}^{g} \chi_{m}^{r}$ & $\chi_{e}^{r} \chi_{m}^{r}$ & $\chi_m^r$ & $\chi_e^b$ & $\chi_{e}^{b} \chi_{m}^{b}$ & $\chi_{e}^{b} \chi_{m}^{g}$ & $\chi_e^g$ & $\chi_{e}^{r} \chi_{m}^{b}$ & $\chi_m^g$ & $\chi_{e}^{g} \chi_{m}^{b}$ & $\chi_e^r$ & $\chi_{e}^{g} \chi_{m}^{g}$ & ${\bf 1}$ & $\chi_m^b$ & $\chi_{e}^{r} \chi_{m}^{g}$ \\
$\chi_{e}^{b} \chi_{m}^{g}$ &  $\chi_{e}^{b} \chi_{m}^{g}$ & $\chi_{e}^{g} \chi_{m}^{g}$ & $\chi_{e}^{r} \chi_{m}^{g}$ & $\chi_m^g$ & $\chi_{e}^{b} \chi_{m}^{b}$ & $\chi_e^b$ & $\chi_{e}^{b} \chi_{m}^{r}$ & $\chi_{e}^{g} \chi_{m}^{b}$ & $\chi_e^r$ & $\chi_m^r$ & $\chi_e^g$ & $\chi_{e}^{r} \chi_{m}^{b}$ & $\chi_{e}^{g} \chi_{m}^{r}$ & $\chi_m^b$ & ${\bf 1}$ & $\chi_{e}^{r} \chi_{m}^{r}$ \\
$\chi_{e}^{g} \chi_{m}^{b}$ &  $\chi_{e}^{g} \chi_{m}^{b}$ & $\chi_{e}^{b} \chi_{m}^{b}$ & $\chi_m^b$ & $\chi_{e}^{r} \chi_{m}^{b}$ & $\chi_{e}^{g} \chi_{m}^{g}$ & $\chi_{e}^{g} \chi_{m}^{r}$ & $\chi_e^g$ & $\chi_{e}^{b} \chi_{m}^{g}$ & $\chi_m^r$ & $\chi_e^r$ & $\chi_{e}^{b} \chi_{m}^{r}$ & $\chi_m^g$ & $\chi_e^b$ & $\chi_{e}^{r} \chi_{m}^{g}$ & $\chi_{e}^{r} \chi_{m}^{r}$ & ${\bf 1}$ \\
\end{tabular}
\label{tab:fusion_table}
\end{table}
\end{center}

\section{Quick guide to entanglement entropy}
\label{AP:EE}

Given a normalized wave-function $\ket{\phi}$ and a partition of the system into subsystems $A$ and $B$, the reduced density matrix of subsystem $A$ is given by \cite{Chuang}
\be
\rho_A={\rm Tr}_B \ket{\phi}\bra{\phi}.
\ee
The Renyi entanglement entropy of the system is then defined as
\be\label{eq:renyi_EE}
S_n=\frac{1}{1-n} \log_2 ({\rm Tr} \rho_A^n),
\ee
where $n$ is the Renyi index. For the limit $n\longrightarrow 1$, Eq.~(\ref{eq:renyi_EE}) is reduced to the von Neumann entropy 
\be\label{eq:VN_EE}
S^{\rm vN}=-{\rm Tr} (\rho_A \log_2 \rho_A).
\ee

For 2D gaped phases with topological order, the Renyi entanglement entropy of a contractible region $A$ with smooth boundary of length $L$ reads \cite{K-P-TEE,L-W-TEE}
\be\label{eq:S_n}
S_n=\alpha L-\gamma+ \mathcal{O} (\frac{1}{L}),
\ee
where the first term arises from the local contribution of the EE which is non-universal and scales linearly with the boundary and the second term, $\gamma$, is a universal
constant which has a topological nature and is called the topological entanglement entropy. 
The explicit value of $\gamma$ depends on the topological nature of the model and is given by
\be\label{eq:gamma_dim}
\gamma=\log_2 \mathcal{D}, \quad \mathcal{D}=\sqrt{\sum_q d_q^2},
\ee
where $\mathcal{D}$ is the total quantum dimension of the model. Here the sum runs over all superselection sectors of the model containing 
a quasiparticle with charge $q$ and $d_q$ is the quantum dimension of the QPs. In Abelian theories,
the quantum dimension $d_q=1$ for all of the QPs. 

The ground state of the topological spin liquids soch as color code and toric code is a uniform superposition of all configurations of closed loops.
The basic intuition implies that the topological contribution of EE, $\gamma$, actually stems from 
these closed strings, which have already been the signatures of topological order. Kitaev and Preskill \cite{K-P-TEE} proved this statement by showing that the 
TEE of disk shape regions with contractible boundaries, such as those in Fig.~\ref{fig:k_p_partitioning} is given by
\be\label{eq:KP-TEE}
-\gamma=S_{ABC}-S_{AB}-S_{AC}-S_{BC}+S_{A}+S_{B}+S_{C},
\ee
where $S$ is the von Neumann entanglement entropy of each subregion. The subregions are strategically
chosen to ensure that local contributions of the entropies are canceled out and what is left is of topological nature i.e. a closed string.
Eq.~ (\ref{eq:KP-TEE}) is particularly suitable for calculating the TEE, numerically.

\section{Extracting the $\Um$ and $\Sm$ modular matrices for a lattice with $2\pi/3$ symmetry}
\label{AP:modular_matrices}

Denoting $V_{ij}=\bra{\MES^{y}_i} R_{2\pi/3} \ket{\MES^{y}_j}$ we can extract the $D$, $\Um$, and $\Sm$-Matrix of the TCC, using the procedure presented in Ref.~\cite{Vidal}.
Having in mind the definition of $\Um$, $D$ and $\Sm$ matrices, the matrix elements of the scalar product of the MESs are given by  
\be
V_{ij}= e^{-i(\phi_i-\phi_j)} \Um_{ii} \Sm_{ij}.
\ee
Here, $\Um_{ii}=\theta_i$ where $\theta_i$ is the twist angle of the anyon type $i$. 
Knowing that the the identity particle has trivial self-statistics, $\theta_1=1$, the mutual-statistics of the identity particle is given by 
$\Sm_{i1}=\Sm_{1i}=\frac{d_i}{\mathcal{D}}$. The relative phases of the MESs can therefore be determined from
the $V_{1j}$ elements of the scaler product matrix
\be
V_{1j}=\Um_{11} \Sm_{1j} e^{-i(\phi_1-\phi_j)}.
\ee
The elements of the $\Um$-matrix are further determined from the $V_{j1}$ elements i.e.
\be
\Um_{jj}=\frac{V_{j1}}{\Sm_{j1}}  e^{-i(\phi_1-\phi_j)},
\ee
which correspond to angle twist of each QP.
Eventually, the rest of the elements of the $\Sm$-matrix extracted from the following relation
\be\label{eq:SmatElement}
\Sm_{ij}= \frac{V_{ij}}{\Um_{ii}} e^{-i(\phi_j-\phi_i)}.
\ee
Using Eq.~(\ref{eq:SmatElement}) and the symmetric property of the $\Sm$-Matrix, the whole elements of the modular matrices are readily determined: 
{\scriptsize \be\label{eq:S_U_Matrix}
\Um=
\left(
\begin{array}{cccccccccccccccc}
 1 &  &  &  &  &  &  &  &  &  &  &  &  &  &  &  \\
  & 1 &  &  &  &  &  &  &  &  &  &  &  &  &  &  \\
  &  & 1 &  &  &  &  &  &  &  &  &  &  &  &  &  \\
  &  &  & 1 &  &  &  &  &  &  &  &  &  &  &  &  \\
  &  &  &  & 1 &  &  &  &  &  &  &  &  &  &  &  \\
  &  &  &  &  & 1 &  &  &  &  &  &  &  &  &  &  \\
  &  &  &  &  &  & 1 &  &  &  &  &  &  &  &  &  \\
  &  &  &  &  &  &  & 1 &  &  &  &  &  &  &  &  \\
  &  &  &  &  &  &  &  & 1 &  &  &  &  &  &  &  \\
  &  &  &  &  &  &  &  &  & 1 &  &  &  &  &  &  \\
  &  &  &  &  &  &  &  &  &  & -1 &  &  &  &  &  \\
  &  &  &  &  &  &  &  &  &  &  & -1 &  &  &  &  \\
  &  &  &  &  &  &  &  &  &  &  &  & -1 &  &  &  \\
  &  &  &  &  &  &  &  &  &  &  &  &  & -1 &  &  \\
  &  &  &  &  &  &  &  &  &  &  &  &  &  & -1 &  \\
  &  &  &  &  &  &  &  &  &  &  &  &  &  &  & -1 \\
\end{array}
\right), \quad 
\Sm=
\frac{1}{4} \left(
\begin{array}{cccccccccccccccc}
 1 & 1 & 1 & 1 & 1 & 1 & 1 & 1 & 1 & 1 & 1 & 1 & 1 & 1 & 1 & 1 \\
 1 & 1 & 1 & 1 & 1 & -1 & -1 & 1 & -1 & -1 & -1 & 1 & -1 & 1 & -1 & -1 \\
 1 & 1 & 1 & 1 & -1 & 1 & -1 & -1 & 1 & -1 & 1 & -1 & -1 & -1 & 1 & -1 \\
 1 & 1 & 1 & 1 & -1 & -1 & 1 & -1 & -1 & 1 & -1 & -1 & 1 & -1 & -1 & 1 \\
 1 & 1 & -1 & -1 & 1 & 1 & 1 & 1 & -1 & -1 & 1 & -1 & 1 & -1 & -1 & -1 \\
 1 & -1 & 1 & -1 & 1 & 1 & 1 & -1 & 1 & -1 & -1 & 1 & -1 & -1 & -1 & 1 \\
 1 & -1 & -1 & 1 & 1 & 1 & 1 & -1 & -1 & 1 & -1 & -1 & -1 & 1 & 1 & -1 \\
 1 & 1 & -1 & -1 & 1 & -1 & -1 & 1 & 1 & 1 & -1 & -1 & -1 & -1 & 1 & 1 \\
 1 & -1 & 1 & -1 & -1 & 1 & -1 & 1 & 1 & 1 & -1 & -1 & 1 & 1 & -1 & -1 \\
 1 & -1 & -1 & 1 & -1 & -1 & 1 & 1 & 1 & 1 & 1 & 1 & -1 & -1 & -1 & -1 \\
 1 & -1 & 1 & -1 & 1 & -1 & -1 & -1 & -1 & 1 & 1 & 1 & 1 & -1 & 1 & -1 \\
 1 & 1 & -1 & -1 & -1 & 1 & -1 & -1 & -1 & 1 & 1 & 1 & -1 & 1 & -1 & 1 \\
 1 & -1 & -1 & 1 & 1 & -1 & -1 & -1 & 1 & -1 & 1 & -1 & 1 & 1 & -1 & 1 \\
 1 & 1 & -1 & -1 & -1 & -1 & 1 & -1 & 1 & -1 & -1 & 1 & 1 & 1 & 1 & -1 \\
 1 & -1 & 1 & -1 & -1 & -1 & 1 & 1 & -1 & -1 & 1 & -1 & -1 & 1 & 1 & 1 \\
 1 & -1 & -1 & 1 & -1 & 1 & -1 & 1 & -1 & -1 & -1 & 1 & 1 & -1 & 1 & 1 \\
\end{array}
\right)
\ee}

%


\begin{thebibliography}
\bibitem{note1}
This can be confirmed visually by painting the QPs and their attached strings on a 2-colex and then exchanging their position according to the coloring 
rules of the TCC. One can see that the strings attached to boson don not anti-commute when the QPs are exchanges. However, the attached strings to fermions
cross each other at a certain vertex and the wave function picks up a $-1$
phase, resembling the phase factor that two fermion would pick when they are exchanged.
\end{thebibliography}
 \end{document}